\documentclass[12pt,a4paper]{iopart}
\pdfoutput=1 
\usepackage{graphicx}
\usepackage{amsfonts}%
\usepackage[utf8]{inputenc}
\usepackage[english]{babel} 
\usepackage[T1]{fontenc}
\usepackage{color}
\usepackage[pdftex,colorlinks]{hyperref}

\begin{document}

\title[An investigation on the high density transition of the SOL transport in AUG]{An experimental investigation on the high density transition of the Scrape-off Layer transport in ASDEX Upgrade}
\author{D. Carralero $^{1}$, G. Birkenmeier$^{1}$, H.W. Müller$^{1}$, P. Manz $^{1,2}$, P. deMarne$^{1}$, S.H. Müller$^3$, F. Reimold$^1$, U. Stroth$^{1}$, M. Wischmeier$^{1}$, E. Wolfrum$^{1}$, and the ASDEX Upgrade team.}
\address{$^1$ Max Planck Institute for Plasma Physics, Boltzmannstr. 2, 85748 Garching, Germany}
\address{$^2$ Physik-Department E28, Technische Universität München, Garching, Germany}
\address{$^3$ Center for Energy Research and Center for Momentum Transport and Flow Organization, University of California, San Diego, La Jolla, USA.}
\ead{daniel.carralero@ipp.mpg.de}

\begin{abstract}
A multidiagnostic approach, utilizing Langmuir probes in the midplane, X-point and divertor walls, along with Lithium beam and infrared measurements is employed to evaluate the evolution of the Scrape-off Layer (SOL) of ASDEX Upgrade across the L-mode density transition leading to the formation of a density shoulder. The flattening of the SOL density profiles is linked to a regime change of filaments, which become faster and larger, and to a similar flattening of the $q_{\parallel}$ profile. This transition is related to the beginning of outer divertor detachment and leads to the onset of a velocity shear layer in the SOL. Experimental measurements are in good agreement with several filament models which describe the process as a transition from conduction to convection-dominated SOL perpendicular transport caused by an increase of parallel collisionality. These results could be of great relevance since both ITER and DEMO will feature detached divertors and densities largely over the transition values, and might therefore exhibit convective transport levels different to those observed typically in present-day devices.\\
\end{abstract}

\maketitle

\section{Introduction}\label{intro}

One of the key open questions regarding the operation of a tokamak fusion reactor is how the energy and particle losses from the confined plasma will distribute on the different plasma facing components (PFC), as it will determine their heat loads, sputtering levels, tritium retention and lifetime. This depends on the transport processes regulating the width of the Scrape-off Layer (SOL). In particular, the distribution of heat and particle fluxes between divertor and main chamber wall depends on the balance between parallel and perpendicular transport in the SOL. After the leading role of diffusion had been called into question \cite{Endler95, Umansky98}, the outward propagation of intermittent convective structures known as ``filaments'' or ``blobs'' has become widely accepted as the dominant mechanism for cross-field transport on the low-field side of tokamaks.  Plasma filaments have been modeled as poloidally localized, coherent structures of high pressure that extend along the magnetic field lines within a background plasma of lower pressure. Curvature-driven instabilities (interchange/ballooning) compete with fluctuating parallel currents to give rise to a poloidal polarization \cite{Endler95}. This polarization causes an $E\times B$ drift which propels the filament in the perpendicular direction. During the last two decades, filaments have been the subject of intense experimental and theoretical study \cite{Zweben07, Dippolito10}. They have been observed to cause cross-field convection to dominate the SOL in many major tokamaks including ASDEX \cite{Endler95}, JT-60U \cite{Asakura97}, Alcator C-Mod \cite{Umansky98,Labombard01}, DIII-D \cite{Bohedo01}, ASDEX Upgrade \cite{Neuhauser02} and TCV \cite{Garcia06b}.\\

One of the best documented experimental features of filamentary transport in the SOL of tokamaks is the L-mode ``high density transition'' (HDT) taking place after a certain density around $f_{GW} \simeq 0.5$, where $f_{GW}=\bar{n}_c/n_{GW}$ is the ratio between the plasma core line density, $\bar{n}_c$, and the Greenwald density, $n_{GW} = I_p/\pi a^2$ ($I_p$ and $a$ are the plasma current and minor radius). After such density is surpassed in the main plasma, the density profile in the SOL changes, flattening its gradient and giving rise to a ``shoulder''. Although this transition had already been reported in other machines \cite{McCormick92,Rohde96,Asakura97,Pitts01}, it was first linked to enhanced perpendicular convective transport (which ``could be dominated by large-scale convection cells'') in Alcator C-Mod \cite{Labombard01}: For $f_{GW} < 0.5$, the SOL displayed two regions, with a steep gradient in the near SOL, followed by a flat gradient towards the wall. After a detailed analysis of SOL transport, LaBombard et al. \cite{Labombard01} conclude that in the first region parallel conduction dominates over perpendicular transport while the far SOL is dominated by perpendicular convection. After $f_{GW} \simeq 0.5$ is surpassed, the flat gradient region extends from the wall to the separatrix and convection dominates the whole SOL. In TCV \cite{Garcia07}, a similar change of the density profile is observed: at fGW~0.25 there are two distinct decay lengths, while at fGW~0.6 there is only one. The high density profile is in very good agreement with ESEL simulations, which describe the transport in the SOL as the result of interchange turbulence leading to the ejection of filamentary structures \cite{Garcia06b}. In DIII-D \cite{Rudakov05} and ASDEX Upgrade (AUG) \cite{Rohde96}, a ``shoulder'' appears in the density profile after the HDT. In DIII-D, the transition is found at $f_{GW} = 0.5$, and leads to a clear decrease in the far SOL gradient, although it does not affect the near SOL. It is also explained as an increase in the ratio of perpendicular to parallel transport, $\Gamma_\bot/\Gamma_\parallel$, due to an enhanced convection. In AUG, the HDT is found \cite{Neuhauser02} at some density between $f_{GW}=0.33$ and $0.62$, although the two slopes can't be distinguished clearly at lower densities. The transition is explained as an increase in filamentary transport. Finally, a similar transition is observed in JT-60U \cite{Asakura97} at a slightly higher density ($f_{GW}=0.55$). \\

This transition in SOL transport has been proposed as one of the key elements in the density limit: Indeed, in Alcator C-Mod the density limit disruption has been observed when the combination of radiative and convective losses exceeded the input power \cite{Labombard01}. This happened when the flat convective region of the SOL reached the separatrix and filaments began forming in the confined region \cite{Terry05}. \\

Besides, part of the literature reporting the HDT investigates the relation between midplane transport and divertor evolution: Experiments carried out in ASDEX \cite{McCormick92} revealed that the divertor strike line was broadened when the divertor electron temperature dropped as a result of increased perpendicular losses. Early work in AUG \cite{Rohde96} suggested that the detachment of the divertor takes place around the same density at which the shoulder appears in the midplane. Finally, the change of SOL profiles in JT-60U is related to the divertor detachment \cite{Asakura97} (in this case, associated to the onset of a MARFE near the X-point). The drop in $T_e$ leads to an increase of the density e-folding length, $\lambda_n$, although in this case it is interpreted as a drop in $\Gamma_\parallel$, and not as an enhancement of convection. Interestingly, the reported detachment of the JT-60U divertor takes place at a higher density fraction (around $f_{GW} \simeq 0.6$) than in other devices, due to its (then) open divertor configuration \cite{Asakura99}. This difference is also observed for the onset of the midplane density shoulder, suggesting a clear link between them.\\

In all these cases, the increase of the far SOL e-folding length appears to be caused by an increased perpendicular to parallel transport ratio, typically linked to an enhanced convection of filaments. One likely explanation for this is the increase of ion-electron collisions as the result of the increase of SOL density and the associated reduction in temperature: In Endler's model, the rise of the Spitzer resistivity along the field lines causes the growth rate of the filament instability, to increase as $\gamma \propto T_e^{-3/2}$ as $T_e$ decreases \cite{Endler95}. According to the mixing length estimate, this would mean an increase in the perpendicular convective transport. Also, the results from simulations by Garcia et al. \cite{Garcia06}, which are in agreement with experimental results from TCV \cite{Garcia06b}, reveal that filaments propagating in the sheath connected regime have lower radial velocities than those disconnected from the sheath due to an increase in collisionality. Early models \cite{Krash01} only considered sheath connected filaments with a parallel structure based on the ``simple SOL'' \cite{Stangeby}, where the filament circuit was closed through the wall and the main potential drop was at the sheath. However, successive refinements \cite{Krash07, Manz13} included alternative closure schemes which defined different scalings of filament properties such as size, amplitude and velocity.  These closure schemes are determined by SOL characteristics such as X-point fanning, finite $\beta$, parallel advection of vorticity, resistivity or $T_e/T_i$. In this context, Myra et al. \cite{Myra06} developed a model where the midplane becomes disconnected from the divertor for large collisionalities. In their invariant scaling, this transition is determined by the effective collisionality parameter

\begin{equation}
\Lambda=\frac{L_c/c_s}{1/\nu_{ei}}\frac{\Omega_i}{\Omega_e},  \label{eq1}
\end{equation}

where $L_c$ is the connection length, $c_s$ the sound speed, $\nu_{ei}$ the electron-ion collision frequency and $\Omega_\alpha$ the gyrofrequency of species $\alpha$. $\Lambda$ can be seen as the ratio between the parallel characteristic length and the ion-electron mean free path. According to the model, for $\Lambda<1$, the midplane is connected to the wall, but for $\Lambda>1$ filaments undergo a regime transition which strongly increases their radial convection. Later work \cite{Myra07} has generalized this single filament-effect to the transport of a turbulent SOL. This mechanism has been invoked \cite{DIppolito06} to explain the HDT in the Alcator C-Mod experiment \cite{Labombard01} and its relation to the density limit. According to this interpretation, convection becomes a feedbacked channel for energy loss: as density increases over a certain threshold $\Lambda\propto n_eT_e^{-3/2}$ is increased, enhancing convection and resulting in less energy transported in the parallel direction. As a consequence, $T_e$ would drop downstream, thus further increasing $\Lambda$. This would explain the formation of the convective profile ``shoulder'' over a certain density threshold which moves closer to the separatrix as plasma density is increased and the divertor effects mentioned before \cite{McCormick92, Rohde96, Asakura97}: An increase of $\Lambda$ would lead to a change in the $\Gamma_\bot/\Gamma_\parallel$ balance, a wider SOL and, finally, to divertor detachment.\\

Although a general picture of filamentary transport and its relation to the transition in the SOL density profile is well established, a more quantitative description is still needed in order to understand the links to the density limit and connect midplane and divertor detachment physics and eventually achieve predictive capabilities. In this sense, the HDT is the ideal testbed to benchmark filament models with the potential to predict the $\Gamma_\bot/\Gamma_\parallel$ ratio into a wider range of plasma conditions, including H-mode operation. A good example of this is the degradation of the confinement observed at high density H-mode operation in AUG \cite{Bernert13}, where a similar change in the structure of the SOL is observed as density is increased, and filamentary transport might play a role as an energy loss mechanism leading to the H-L back transition and eventual disruption of the plasma. In this context, the purpose of this work is twofold: On a first level, the HDT in AUG is investigated with an improved set of diagnostics in order to determine with precision its density threshold and to measure the associated changes in the SOL transport. On a second level, we use the measurements to test the validity of the mechanisms discussed above, and their effect (or lack thereof) on the rest of the SOL. With this purpose, we provide a comprehensive experimental coverage of the SOL evolution across the transition in order to analyze simultaneously the changes in midplane perpendicular transport, turbulence characteristics, collisionality parameter $\Lambda$ and the effect on the divertor to obtain a consistent picture of all the related phenomena.\\

The paper is organized as follows: In section \ref{Exp} we describe the experimental setup used in AUG, along with the diagnostics employed and the characteristics of the plasma discharges. Next, the methodology for the analysis of turbulence is described in section \ref{anatech}. Experimental results are presented in section \ref{ExpRes}. Then, all results are discussed and compared to the previously presented models in section \ref{Discussion}. Finally, we present our conclusions and examine future lines of work in section \ref{conclusion}.\\

\section{Experimental Setup}\label{Exp}

\begin{figure}
	\centering
		\includegraphics[width=\linewidth]{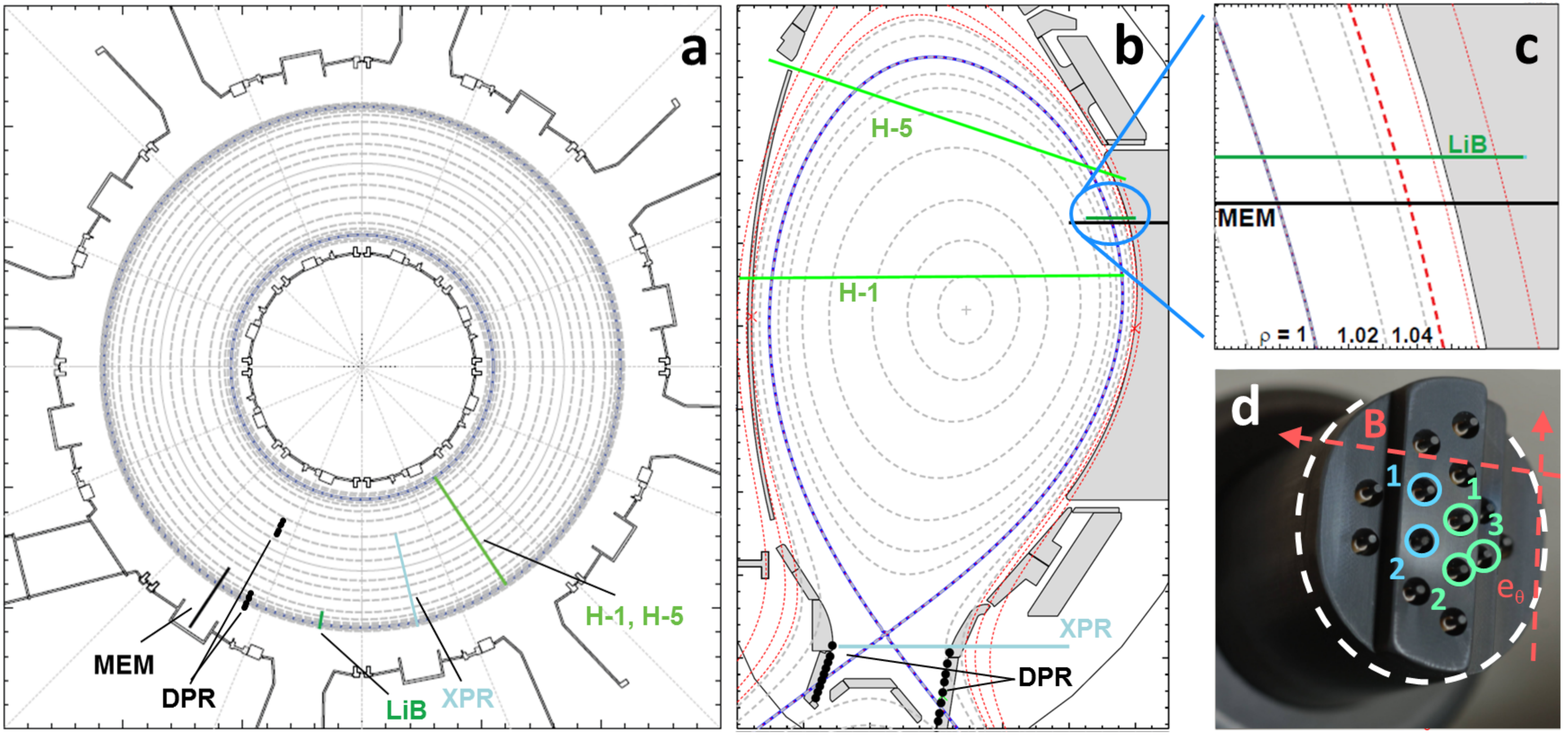}
	\caption{\textit{General experimental layout. (a) and (b) are respectively toroidal and poloidal sections of a typical discharge. Separatrix is indicated as a dashed blue line. Grey dashed curves represent flux surfaces. Surfaces intersecting the limiters are indicated as red dashed curves. Positions of main diagnostics are indicated: Midplane manipulator (MEM),  X-point manipulator (XPR), divertor fixed probes (DPR), Lithium beam (LiB) and core and edge interferometer lines (H-1 and H-5, respectively). (c) shows the detailed structure of the SOL in the measurement region of MEM and LiB, including the first limiter shadow, displayed as a thick dashed red line. (d) shows the 14 pin probe head installed in the MEM. Parallel and binormal directions are indicated in red. Blue and green circles indicate pins measuring $\phi_f$ and $I_{sat}$, respectively. The last cylindrical section is indicated as a dashed white curve.}}
	\label{fig:diag}
\end{figure}

The experiments reported in the present work were carried out in the ASDEX Upgrade tokamak. AUG is a medium sized divertor tokamak with major and minor radii of $R = 1.65$ m and $a = 0.5$ m, respectively. It is operated with full tungsten coated walls \cite{Kallenbach11}. Fig. \ref{fig:diag} shows the general experimental layout: In Fig. \ref{fig:diag}(a) and (b), the typical magnetic geometry used during the experiments is displayed inside the vessel. As can be seen, it is a single null divertor configuration in which the low field side is tailored to follow the shape of the outer limiter. In Fig. \ref{fig:diag}(c), the detailed structure of the SOL in the midplane measurement region is displayed using the normalized poloidal magnetic flux coordinate, $\rho$: Up to $45$ mm in front of the separatrix, field lines are connected to the divertor targets. Beyond that point (around $\rho > 1.04$), they become shadowed by the inner limiter and the connection length $L_c$ drops to a few meters.\\

The position of diagnostics is also displayed in Fig. \ref{fig:diag}. The main diagnostic is the Midplane Manipulator (MEM), equipped with a carbon probe head designed to characterize turbulence, featuring 14 carbon pins distributed among three terraces at different radial positions and different arrays aligned in the binormal direction $\textbf{e}_\theta=\textbf{e}_r\times\textbf{e}_\parallel$. These pins can measure either ion saturation current, $I_{sat}$, or floating potential, $\phi_f$, at acquisition rates of 2 MHz. The probe head, described in detail in Ref. \cite{Nold10}, is displayed in Fig. \ref{fig:diag}(d), including the pin layout. Each pin has a diameter of $0.9$ mm, a length of $2$ mm. An effective probe surface of $A_{eff} \simeq 5$ mm$^2$ has been assumed\cite{Nold10}. The binormal separation between pins $1$ and $2$ of both types is $L_\theta = 5.5$ mm, while the terrace of pin 3 is retracted by $L_r = 4$ mm. The orientation of the terraces is such that pin 3 is magnetically connected to the LFS divertor region. The positive binormal direction is defined as the electron-diamagnetic drift velocity $u_{dia,e} = \nabla p_e \times {\bf B}/enB^2$, directed upwards at the low-field side of the torus. The last cylindrical section of the probehead is also indicated, as it will be used as a reference in section \ref{transport}. The probe head is inserted horizontally around $30$ cm above the midplane by the fast reciprocation of the MEM, thus measuring also radial profiles in the far SOL. In order to obtain information on the downstream conditions of the SOL, radial profiles of $I_{sat}$ are measured with a second manipulator situated in the vicinity of the X-point (XPR). The XPR is installed at $z=-0.996$ m and reciprocates horizontally through the divertor entrance, falling below the X point by about 2-5 cm \cite{Tsalas05}. Its probe head is equipped with three graphite pins, measuring $T_e$ (sweeping) and $I_{sat}$. Finally, two arrays of fixed Langmuir triple probes (DPR) are used at low sampling rates ($33$ kHz) to measure $T_e$ and $n_e$ at the inner and outer divertor. These probes, represented as black dots in Fig. \ref{fig:diag}, are flush mounted in the tiles of the inner and outer divertors and have typical dimensions of $25$ mm in the toroidal direction and $4$ mm in the poloidal direction.\\

A Lithium beam (LiB) was used to determine the density profiles in the SOL and edge of the plasma. This diagnostic, which was recently upgraded with a new bundle of lines of sight \cite{Willendorfer13}, observes approximately the same poloidal region as the MEM (although with a toroidal separation of about $25^\circ$), and features high sampling rate (200 kHz) density measurements in the $\rho \in [0.9,1.05]$ range. This range is divided in $26$ radially distributed elliptical channels with a width of $6$ mm in the radial direction and $12$ in the direction perpendicular to the Lithium beam. Averaged line densities in the core and edge of the plasma, $\bar{n}_c$ and $\bar{n}_e$ are measured with the standard interferometer system from AUG (taking a length inside the separatrix of $1$ m). In particular, lines $H-1$ and $H-5$ are used for this, represented in Fig. \ref{fig:diag} as light green lines. Finally, a fast IR camera observes the MEM probe head. Using an absolute calibration of the camera, the temperature on the surface of the probe can thus be derived. After correcting the jitter, $q_{||}$ to the surface of the probe can be calculated from the temperature measurements using the THEODOR code \cite{Herrmann01}. This technique has been successfully used in the past \cite{Herrmann95} to carry out calorimetry measurements. IR measurements of $q_{||}$ were found to be in a good agreement with earlier measurements of the ELM-averaged q|| by a retarding field analyzer probe
installed on MEM \cite{Kocan12ELM}.\\

\begin{figure}[t!]
	\centering
		\includegraphics[width=.85\linewidth]{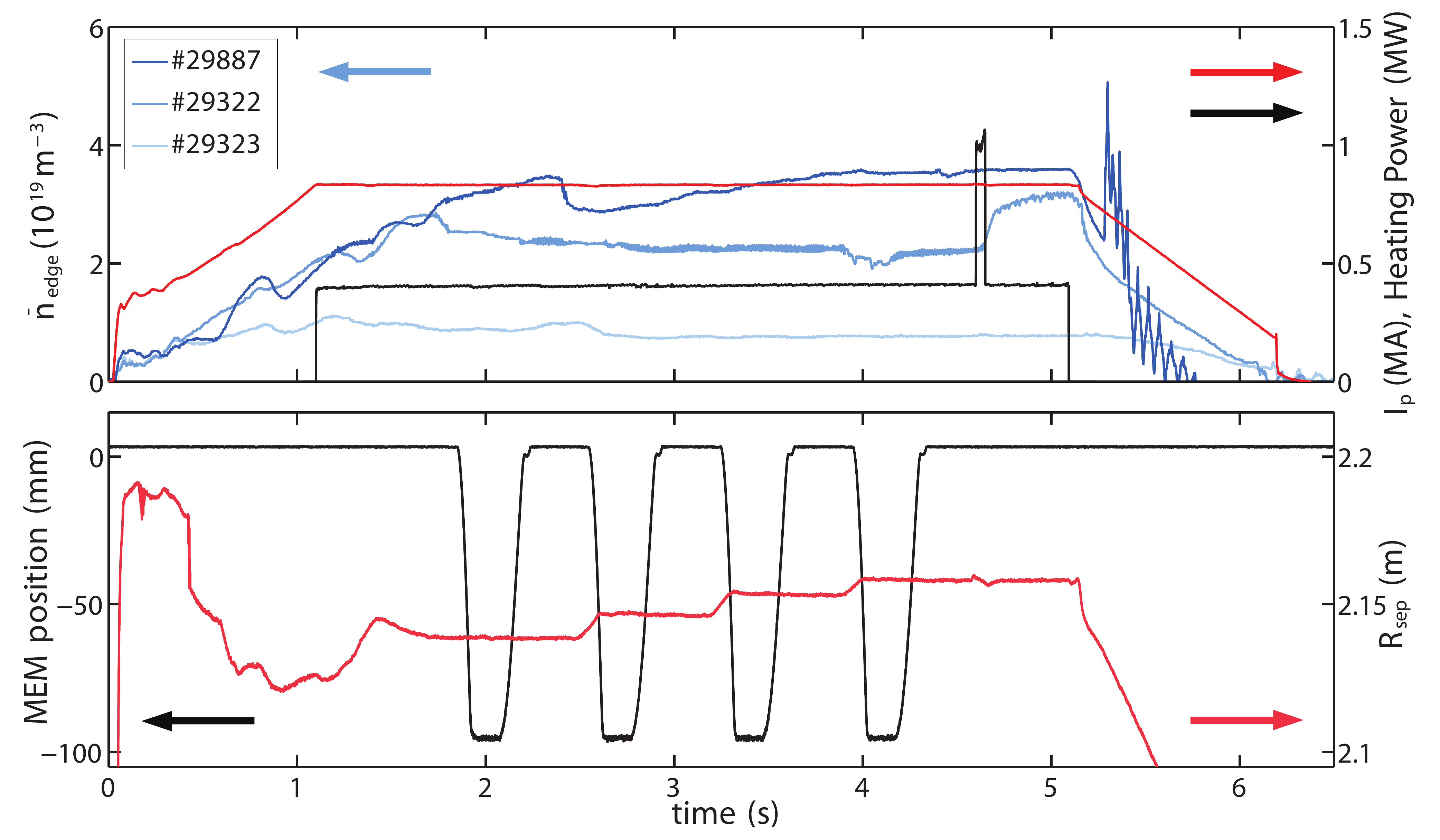}
\caption{\textit{Typical experimental discharge: On top, $I_p$ and external heating power (including ECRH and the NBI blip) are represented as red and black curves, respectively. Edge line densities, $\bar{n}_e$, are represented in different shades of blue from different discharges (from light to dark, \#29323, \#29322 and \#29887). On the bottom, the radial displacement of the probe head and the separatrix at the midplane are represented as black and red curves, respectively.}}\label{fig0}
\end{figure}

For this work, we have aimed at reproducing plasma conditions equivalent to those in which the HDT has been reported in literature: a number of L-mode discharges were carried out with constant magnetic parameters (toroidal magnetic field $B_t = -2.5$ T, safety factor $q_{95}= 5.32$, plasma current $I_p = 800$ kA) and a constant ECRH power of $600$ kW. The Greenwald density was therefore also kept constant at a value of $n_{GW} = 1.02 \cdot 10^{20}$ m$^{-3}$. In this series, different fueling levels were used to raise the density on a shot to shot basis. Values of core and edge line density covered a range of $\bar{n}_c \in [1.3,5.8] \cdot 10^{19}$ m$^{-3}$ and $\bar{n}_e \in [0.75,3.5] \cdot 10^{19}$ m$^{-3}$. These densities correspond to $f_{GW} \in [0.15, 0.6]$, the same range reported in the experiments discussed in the introduction. The relatively low value of heating power was selected to remain far from the L-H transition threshold \cite{Ryter} for all density values, thus excluding the effect of confinement changes and pedestal formation on the observed phenomena.\\

\begin{figure}
	\centering
		\includegraphics[width=.4\linewidth]{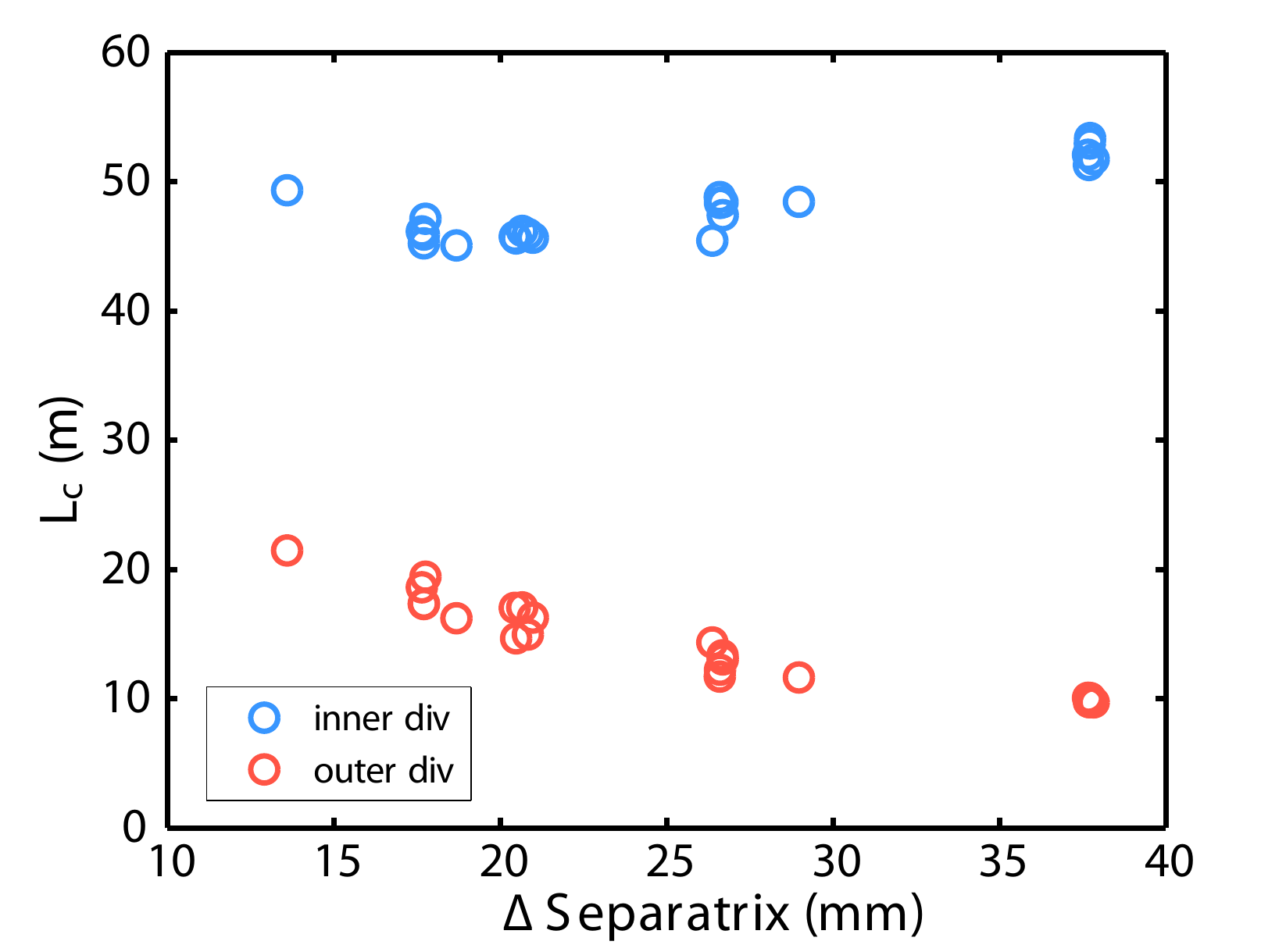}
	\caption{\textit{Connection length, $L_c$, at the MEM position as a function of separation to the separatrix $\Delta$. Blue and red circles represent branches leading to the inner and outer divertor, respectively.  }}\label{fig:Lc}
\end{figure}

A representative plasma discharge is displayed in Fig. \ref{fig0}: The heating power is constant during the whole $I_p$ plateau, only changed by an NBI blip around $t = 4.5$ s. Several density levels are achieved in different discharges. In each discharge, the MEM plunges four times in the SOL, remaining around $100$ ms in its innermost position in order to obtain sufficient statistical data on the turbulent fluctuations. The innermost position is fixed during all the experiments at $R=2.144$ m ($42$ mm in front of the limiter shadow) and $z=0.3$ m. The radius of the separatrix at the outboard mid-plane, $R_{sep}$ is slightly modified by $5$ mm in between plunges in order to achieve different radial distances to the probe. By these means, a range of radial distances between the separatrix and the probe $\Delta \in [15-40]$ mm is achieved. This change in the plasma position leads to different connection lengths at the MEM measurement point. This is shown in Fig. \ref{fig:Lc}, where the length of the magnetic field line from the probe to the inner and outer walls is displayed for the whole range of $\Delta$ values. As can be seen, $L_c$ is always shorter to the outer divertor (at least by a factor of $2$) and despite a moderate dependence on $\Delta$, values of $L_c$ remain always in the same order of magnitude $L_{c,in} \simeq 50$ m, $L_{c,out} \simeq 10-20$ m.\\

\section{Turbulence analysis methodology}\label{anatech}

Fast $\phi_f$ and $I_{sat}$ measurements carried out during the MEM plunges are used to characterize filamentary structures in the SOL. $\phi_f$ and $I_{sat}$ are respectively related to plasma potential $\phi_p$ and  plasma density $n_e$ through the electron temperature, $T_e$. In particular,
\begin{equation}
 \phi_f \simeq \phi_p - 3T_e; \quad I_{sat}= \frac{1}{2}n_eA_{eff}ec_s, \label{eq2}
\end{equation}

where $c_s=\sqrt{(T_e+T_i)/m_i}$ is the isothermal sound speed \cite{Bissell89}. Although no direct $T_e$ or $T_i$ measurements were carried out in the midplane, their approximate values in the SOL are known from previous L-mode experiments carried out in similar conditions. $T_e$ is known to be roughly constant in the SOL region beyond $10$ mm from of the separatrix, with values in the order of $T_e \simeq 15$ eV \cite{Nold10}. Thomson scattering data support this assumption. As for $T_i$, SOL ion temperatures have been measured in the past \cite{KocanRFA} for radial positions in the range $\Delta \simeq 20-40$ mm. Typical values are $T_i\simeq 25$ eV. These measurements correspond to background levels, but not necessarily to filaments, which are typically warmer than the bulk plasma. Indeed, values in the range of $T_i \simeq 80-100$ eV have been measured in filaments (which exhibit a greater dependence on the radial position than the background) \cite{KocanRFA}. Regarding $T_e$ fluctuations, values of up to $30$ eV have been measured during ELMs \cite{HWM10}. L-mode filament deviations from the background electron temperature should be smaller (under $10$ eV), and are therefore neglected. As will be shown in the following section, these approximations allow for a very good agreement between MEM and LiB profile measurements, and are therefore considered to be essentially valid.\\

\begin{figure}
	\centering
		\includegraphics[width=\linewidth]{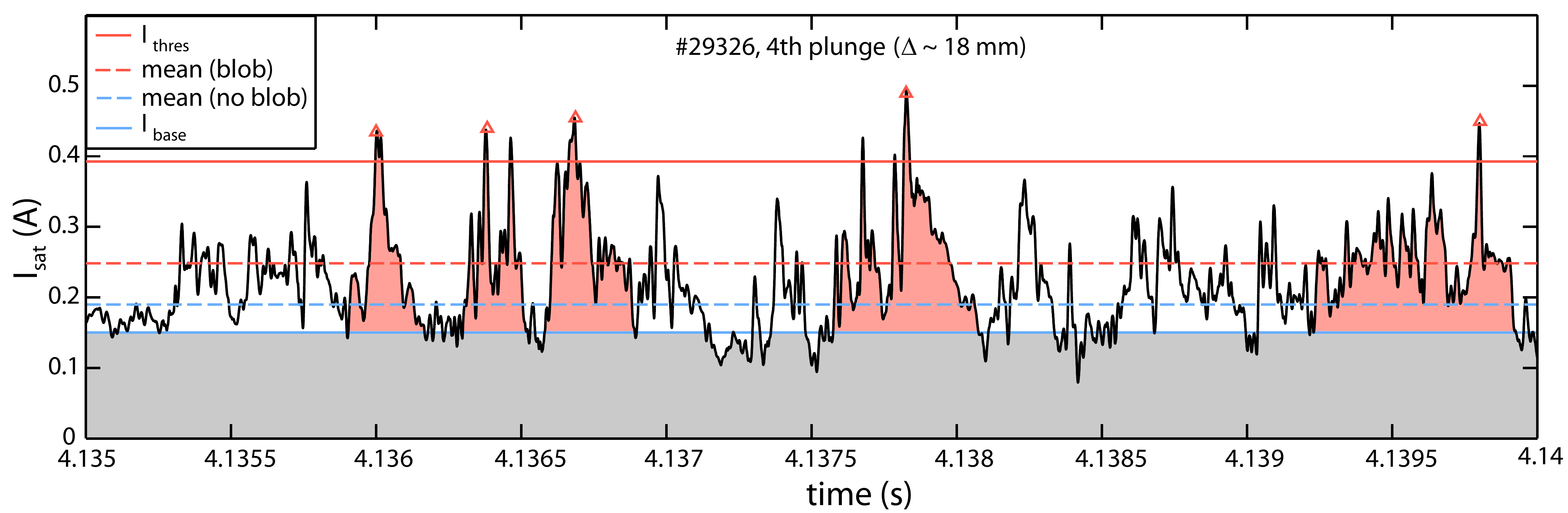}
	\caption{\textit{Filament detection algorithm. Raw data from $I_{sat,1}$ during the fourth plunge of discharge $\#29326$. The red solid line indicates the detection threshold, $I_{thres}$. Detected events are indicated by red triangles. ``Exclusion zone'' around the event is indicated in red. Other lines represent $I_{base}$ (solid blue), average signal without blobs (dashed blue) and average signal during blobs (dashed red).} }	
	\label{fig:fil_lev}
\end{figure}

In Fig. \ref{fig:fil_lev}, a representative example of $I_{sat}$ measurements at the probe pin 1, as defined in Fig. \ref{fig:diag}(d), is displayed for $\Delta = 18$ mm. A first relevant feature is the existence of clear background signal over which large fluctuations can be seen. This generic structure of the signal, is common to all $\bar{n}_e$ and $\Delta$ values and has also been observed by detailed lithium beam studies of SOL turbulence \cite{Birk13}. It indicates that filaments do not propagate in vacuum, but through a plasma medium. Taking this into account, a ``baseline background level'' $I_{base}$ is defined, separating the contribution to the $I_{sat}$ related to the filaments ($I_{sat}>I_{base}$) and to the background ($I_{sat}<I_{base}$). $I_{base}$ is obtained by smoothing the sample using a Savitzky-Golay filter \cite{SavGol} and then calculating the mean value of all the local minima in it. The resulting $I_{base}$ is displayed in Fig. \ref{fig:fil_lev} as a solid blue line, and the ``background'' region is colored in gray. A second feature is the presence of bursty structures, typically designated ``filaments'' or ``blobs''. Following the detection algorithm often found in the literature \cite{Zweben07}, blobs are defined as fluctuations exceeding a threshold, $I_{thres} = \mu+2.5\sigma$, where $\mu$ and $\sigma$ are the mean and standard deviation of the signal. This definition of filament has been selected as it is typically used in large tokamaks \cite{Dippolito10}, and ensures the detection of a sufficient number of filaments (over 50) in all data sets. The threshold is represented in Fig. \ref{fig:fil_lev} as a solid red line. However, one problem of this method is the ambiguity caused by the fine substructure of the fluctuations, since sometimes filaments appear as a train of peaks rather than as a clear maximum. In order to avoid this and prevent multiple detections associated to a single structure, an ``exclusion zone'' in which no new detections are allowed is defined around each point of detection using $I_{base}$ as a second threshold indicating the end of the ``filament structure'': individual filamentary structures comprise the interval around the detection point for which $I_{sat} > I_{base}$. The detection time of the filament is then defined by the local maximum in such an interval. Some examples of filament detections are displayed in Fig. \ref{fig:fil_lev} as triangles. The ``exclusion zone'' around them is colored in red. Once the filaments along the sample have been identified, several levels can be defined: first, the average of $I_{sat}$ of the sample is calculated excluding times in which a filament structure is present. This level is represented as a dashed blue line in the Fig. and is indicative of the mean non-filamentary density. Second, the average of $I_{sat}$ in the sections of the sample containing filament structures is calculated (dashed red line). This indicates the mean density of filaments.\\

\begin{figure}
	\centering
		\includegraphics[width=.4\linewidth]{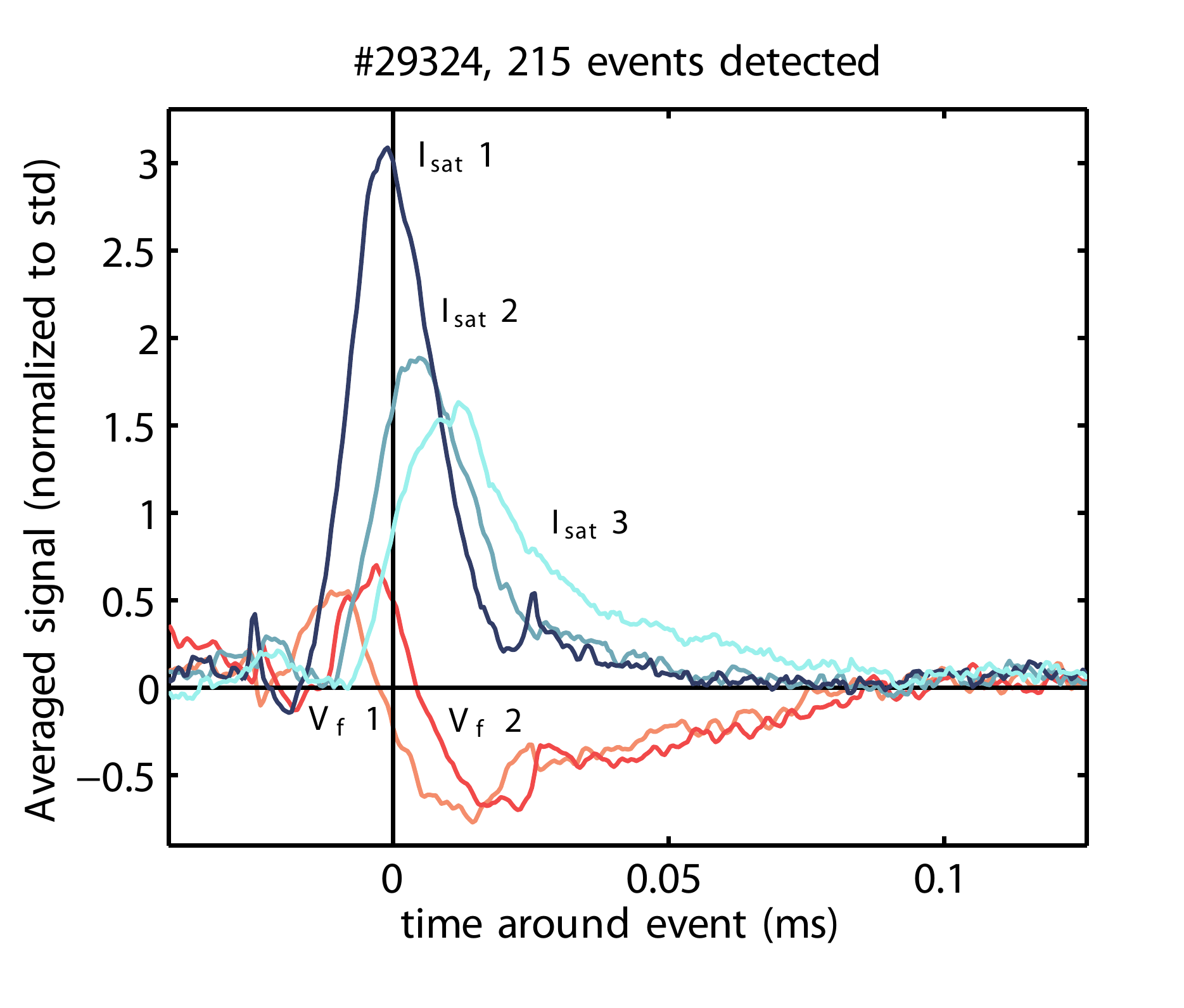}
	\caption{\textit{Conditional averaging example. The signals of the five highlighted pins of Fig. \ref{fig:diag}d are sampled around filament events defined using a $2.5\sigma$ threshold on the $I_{sat,1}$ during the third plunge of shot $\#29324$ ($\Delta = 20.5$ mm, $\bar{n}_e = 1.16 \cdot 10^{19}$ m$^{-3}$).}}
	\label{fig:cond}
\end{figure}

In order to analyze the evolution of filaments, we resort to the conditional averaging \cite{Johnsen} of the multipin probe data to obtain information about their size and propagation: Filament events are detected following the explained procedure on the $I_{sat}$ signal from pin $1$, used as a reference. For each event, the signals of the five highlighted pins of Fig. \ref{fig:diag}(d) are sampled. By averaging over all detected events in a given plunge, the typical filament signature for each pin is obtained for the corresponding $\Delta$ and $\bar{n}_e$. An example of the result of such a procedure averaging is displayed in Fig. \ref{fig:cond}, corresponding to the third plunge of discharge \#29324 ($\Delta = 20.5$ mm, $\bar{n}_e = 1.16 \cdot 10^{19}$ m$^{-3}$). In order to ease the representation, the mean has been subtracted from all signals, which then have been normalized to their standard deviation. An average event can be clearly seen in all five signals: the binormally and radially separated pins $2$ and $3$ show the same structure as pin 1 with a delay caused by the propagation of the filament, while the two $\phi_f$ pins show a similarly delayed structure. Both the tailed shape of the $I_{sat}$ and the bipolar structure of $\phi_f$ are typical filament characteristics, matching reports from many other devices \cite{Dippolito10}. They are also in good agreement with the basic filament model \cite{Garcia06b}, including the dipolar potential structure propelling the filament, which can be seen clearly in the $\phi_f$ signals. Interestingly, for $I_{thres} > \mu+1.5\sigma$, the results of the conditional average do not depend strongly on the selected $I_{thres}$. Of course, this is not the case with the detection rate, which decreases exponentially with $I_{thres}$.\\

Once the characteristic filament structure has been obtained, the average propagation, duration and spatial size of the filaments can be calculated for each available set of $\Delta$ and $\bar{n}_e$. Since filaments dominate the fluctuations, the duration (travel time of the filament over a given pin) obtained in the conditional average does not differ substantially from the autocorrelation time, $t_{AC}$, calculated from the raw signal over the sample using the standard definition found in literature (as the width of the central peak of the autocorrelation function at the $0.5$ level). Since the particular values of conditionally averaged duration depend on the selected threshold, $t_{AC}$ is chosen to define the characteristic duration of structures, as a more standard measure. The propagation of filaments is calculated from the delay between the three $I_{sat}$ pins. In principle, the derivation of average $v_r$ and $v_\theta$ values from three independent pins distributed along the radial and binormal directions (as in this case) is straightforward. However, this task was complicated by the fact that filamentary structures turned out to be substantially larger than the separation between these pins, which causes a strong influence of the filament shape and inclination on the delays. Therefore, some assumptions had to be made in order to include these factors and to avoid unrealistic results \cite{Carralero13}. A more detailed discussion of this issue can be found in the Appendix. Another method frequently found in the literature for the determination of the filament propagation is to calculate the $E\times B$ velocity caused by the potential dipole. This method has not been used in this work since its validity is questionable: recent works demonstrated that under certain conditions the influence of $T_e$ fluctuations on $\phi_f$ in the SOL of AUG is too large to deduce plasma potential fluctuations from it \cite{Nold12}. Finally, once the duration and the propagation velocities have been determined, the characteristic perpendicular size of the filament, $\delta_b$, can be calculated as $\delta_b = v_{\bot}t_{AC}$.\\

\section{Experimental Results}\label{ExpRes}

\subsection{Profiles}\label{prof}

\begin{figure}[t!]
	\centering
		\includegraphics[width=\linewidth]{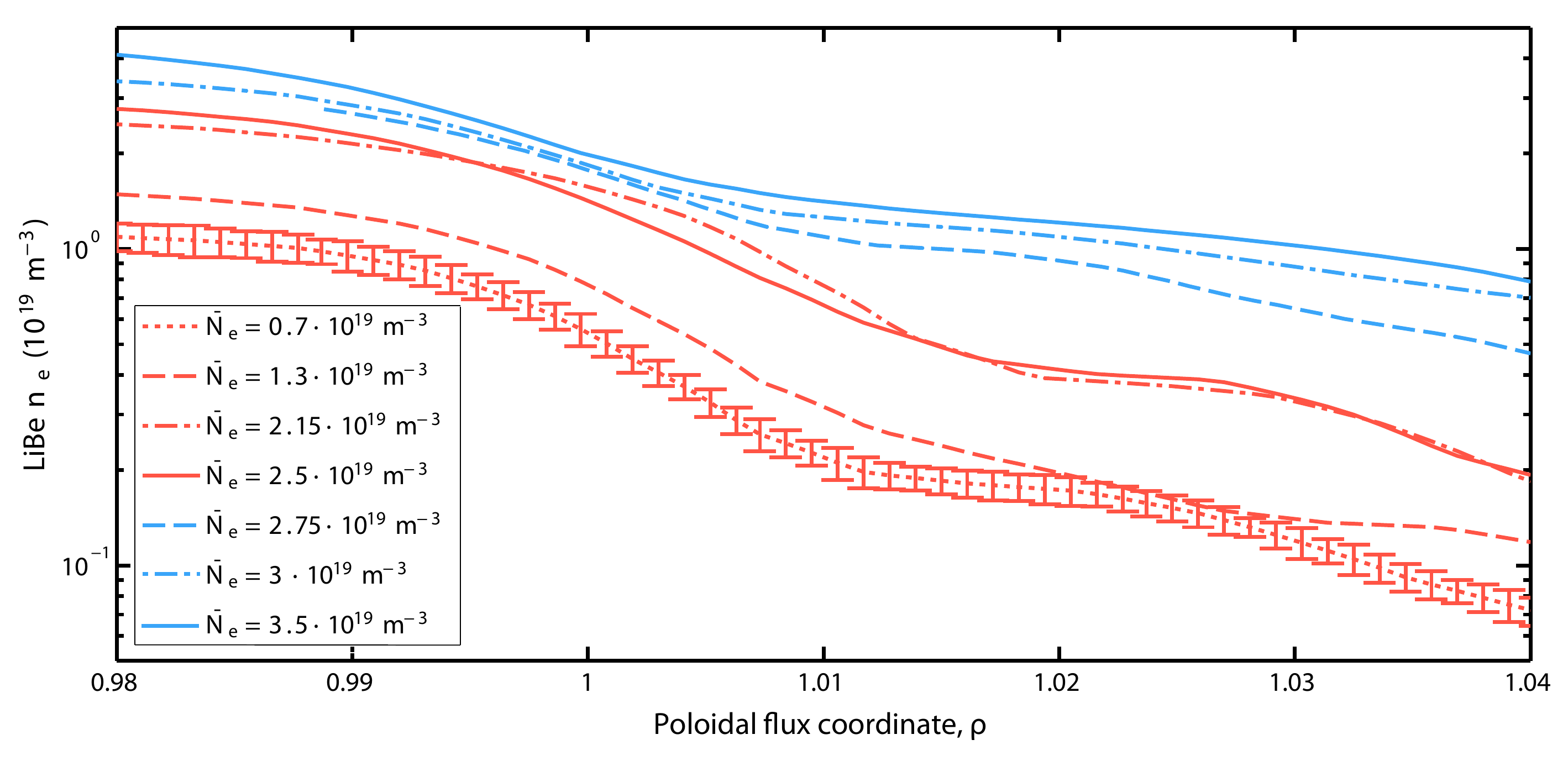}
			\caption{\textit{Evolution of lithium beam radial density profiles for different $\bar{n}_e$ values. Red and blue lines indicate $\bar{n}_e$ under and over $\bar{n}_\textnormal{HDT}= 2.5 \cdot 10^{19}$ m$^{-3}$. Typical error bars are shown in one case as a reference.}}
	\label{fig:figprofiles}
\end{figure}

The evolution of the midplane profile with the increase of line averaged density has been mainly observed with the Lithium beam. In Fig. \ref{fig:figprofiles}, the profiles measured by the LiB are displayed for several representative values of $\bar{n}_e$ as a function of the normalized poloidal magnetic flux coordinate, $\rho$. These profiles are averaged over a 200 ms period centered around the MEM plunge. Since the $\rho$  of the LiB channels changes slightly  during this period, the profiles are splines fitted to the data. Typical error bars are shown in one of the curves. Qualitatively, two kinds of SOL profiles can be distinguished depending on $\bar{n}_e$ being under or over an edge density transition value of $\bar{n}_\textnormal{HDT}= 2.5 \cdot 10^{19}$ m$^{-3}$ (corresponding to a $f_{GW} \simeq 0.45$: The first kind (indicated in the figure by the red color) is characterized by a strong gradient in the vicinity of the separatrix ($\rho\in[1,1.01]$), followed by a flat region towards the far SOL. The second kind lacks such a first gradient and displays a somehow constant e-folding length in the whole SOL, similar to the one observed in the far SOL in the previous case. As a result of this, the density of the far SOL is increased up to the wall, and the gradients remain roughly constant.\\

\begin{figure} [t!]
	\centering
		\includegraphics[width=\linewidth]{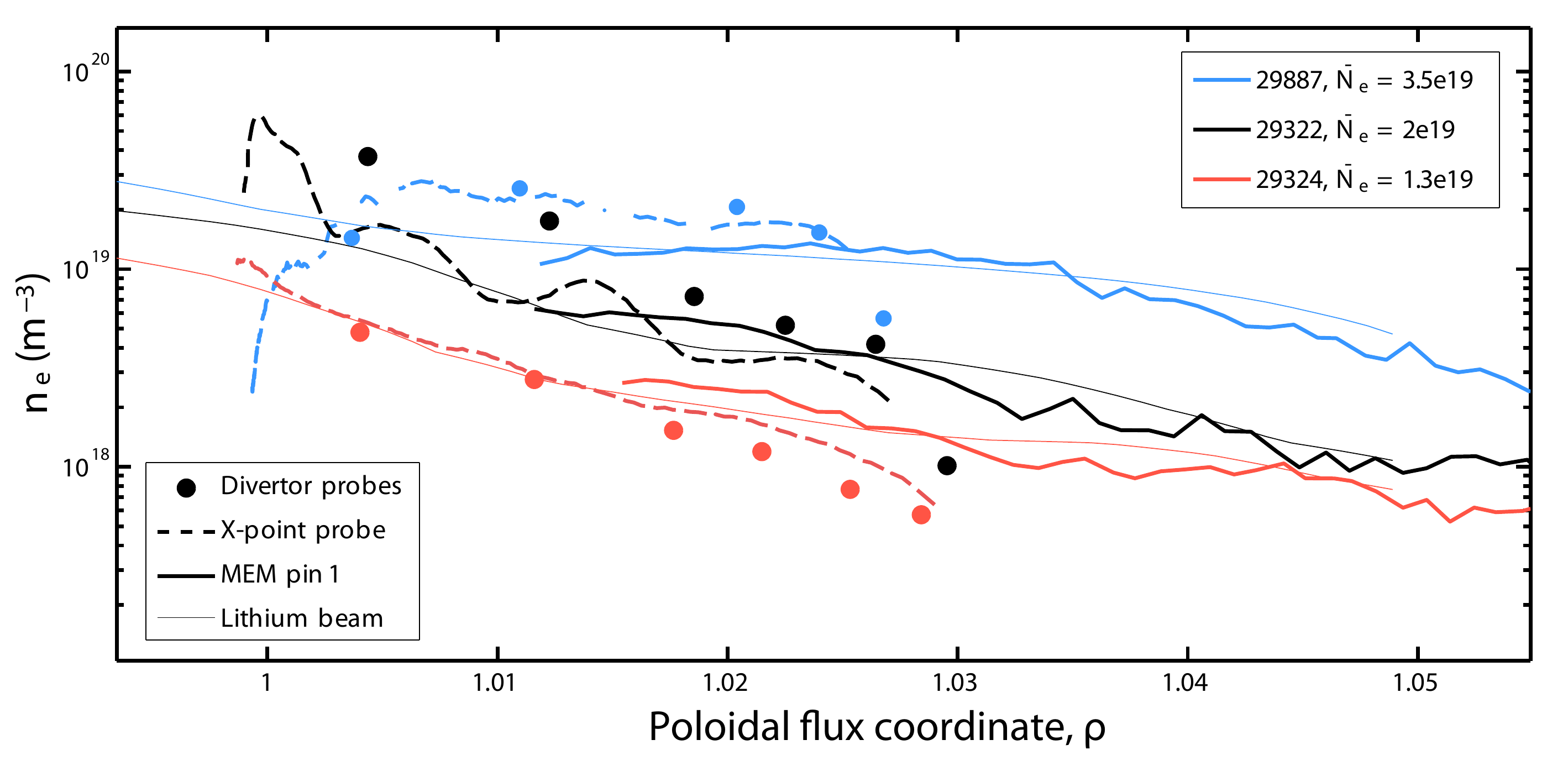}
	\caption{\textit{Comparison of SOL profiles measured by various diagnostics for different values of $\bar{n}_e$. Thin and thick solid lines correspond respectively to LiB and MEM measurements carried out in the midplane. MEM data corresponds to pin $I_{sat,1}$ in Fig. \ref{fig:diag}. Dashed lines correspond to XPR profiles measured below the X-point. Solid circles represent measurements from flush mounted Langmuir probes in the divertor target plates.}}
	\label{fig:profiles_multi}
\end{figure}

Lithium beam data have been complemented by other diagnostics to obtain a comprehensive picture of the evolution of the whole SOL. In the midplane, the fast movement of the MEM allows for the measurement of $I_{sat}$ for a range of $\rho$ values. From it, mean density profiles are derived from $I_{sat,1}$ as explained in the previous section. Downstream, the density profiles below the X point and in front of the outer divertor targets are measured by probes (XPR and DPR). This is displayed in Fig. \ref{fig:profiles_multi} for $\bar{n}_e = 1.3 \cdot 10^{19}$ m$^{-3}$ (below the transition), $\bar{n}_e = 2 \cdot 10^{19}$ m$^{-3}$ (close to the transition) and $\bar{n}_e = 3.5 \cdot 10^{19}$ m$^{-3}$ (above the transition). Since the observed $\rho$ range changes substantially with the variation of $R_{sep}$ between plunges, all three profiles in the figure correspond to the last plunge of each discharge, when the separatrix is closest to the probe final position. Also, due to its perturbative effect on the plasma, the XPR is only plunged once at the end of each discharge. Therefore, only the last plunge of the MEM is directly comparable with XPR data. Since the structure of the presheath is difficult to determine in the divertor region, $n_e$ measurements from the divertor probes are referred to sheath entrance densities (i.e., the $1/2$ factor from equation (\ref{eq2}) is not considered in the divertor). Therefore, the $n_e$ values from the DPR correspond to the position immediately in front of the target. Instead, $n_e$ measured by MEM and XPR refers to unperturbed plasma values at the entrance of the presheath.\\

The first feature seen in Fig. \ref{fig:profiles_multi} is the good agreement between the absolute values of $n_e$ measured by the different diagnostics. In particular, the agreement between the lithium beam and the MEM data is remarkable and indicates that the temperature and effective area approximations involved in the probe signal analysis are reasonable: the difference between the two diagnostics is well within their respective error bars and the precision on the determination of the relative radial positions. A combination of MEM and LiB data reveals how the main changes in the profiles take place in the $\rho \in [1,1.02]$ region. In most of the region measured by the MEM, the profiles increase their density values over the transition, but retain the same gradients. This reinforces the hypothesis of the transition being caused by an increase of the convection in the near SOL. The XPR covers this region and measures a flattening of the profile which follows closely the LiB profile at the midplane. Finally, by comparing the absolute $n_e$ values of the four diagnostics, it is possible to describe the evolution of the parallel structure of the SOL over the transition, relating the changes at the midplane with the evolution of the divertor regime \cite{Stangeby}: prior to the transition, the same density values are found in the midplane, X point and divertor from $\rho = 1$ to $1.02$, indicating the sheath limited (or at least low recycling) regime of the divertor and the validity of the two point model for the SOL. Beyond $\rho = 1.02$, the fall of downstream densities with respect to the midplane doesn't follow any model of the SOL and is probably caused by an error in the calculation of the equilibrium (in the order of 5 mm). Electron temperatures measured at the X-point show a decay from around 40 eV at $\rho =1$ to $20$ eV for $\rho>1.015$. The same trend is followed by the divertor, with slightly lower values (around 75\% of those at the X-point). Also, for $\rho>1.02$, an increase of $T_e$ in the divertor probes over X-point levels might indicate a malfunction of the triple probe sets, which would be over(under)estimating $T_e$($n_e$). Almost at the transition (black color in Fig. \ref{fig:profiles_multi}), profiles at the X-point still remain close to the midplane, but the divertor density increases substantially, indicating the high recycling regime, where the divertor region close to the separatrix begins to cool down. The region beyond $\rho = 1.02$ remains in a low recycling regime. In good agreement with this, $T_e$ in the divertor falls to 50\% of the X-point value ($10$ and $20$ eV, respectively) in the near SOL, and rise to similar values for $\rho > 1.02$. Last, after the transition, the density at the divertor remains higher than the one at the midplane, but its peak moves away from the separatrix, indicating the beginning of the detached regime. XPR measurements follow closely those at the divertor: the radial density profile becomes peaked, and its maximum also moves away from the separatrix. Temperature measurements are consistent with this: after the transition, $T_e$ values fall below $20$ eV in the X-point profile and below $5$ eV for $\rho < 1.025$ at the divertor plates.\\    
 
%
%

\subsection{Changes in turbulence}\label{MEM}

\begin{figure}
	\centering
		\includegraphics[width=\linewidth]{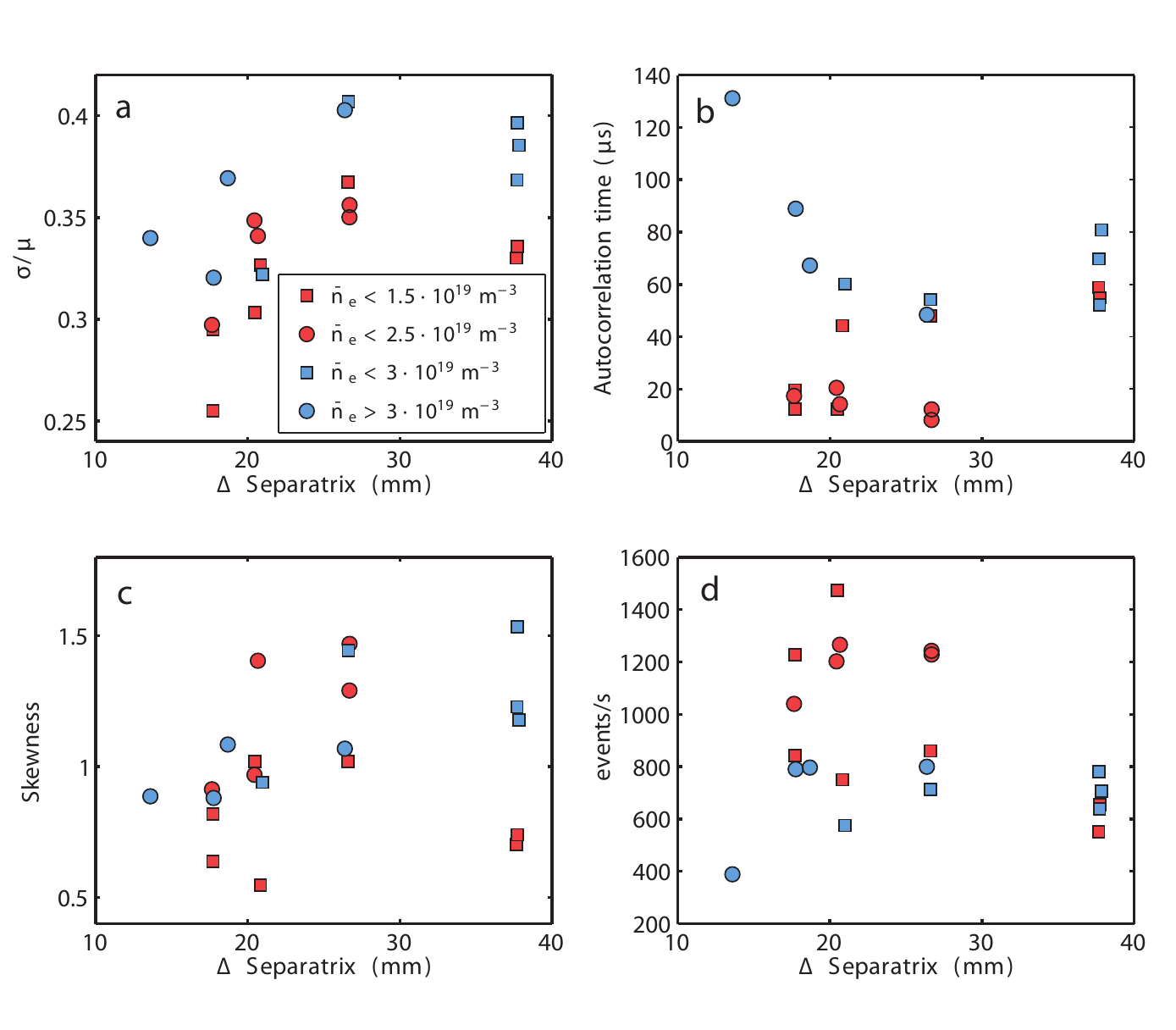}
	\caption{\textit{Radial profiles of statistical properties of midplane turbulence as measured by the MEM Langmuir probes. Different values of $\bar{n}_e$ have been classified into four groups to ease visualization.  As in Fig. \ref{fig:figprofiles}, red and blue points indicate densities below and above the transition. }}
	\label{fig:raw_data}
\end{figure}

In order to carry out a quantitative description of the dependence of fluctuations properties on both $\Delta$ and $\bar{n}_e$, the statistical properties of the $I_{sat}$ signals measured with pin 1 during each plunge have been calculated. This can be seen in Fig. \ref{fig:raw_data}. In order to ease the representation, the various densities corresponding to each plunge have been grouped into four manifolds, two of them below the transition (represented in red) and the other two above (represented in blue). In Fig. \ref{fig:raw_data} (a), the relative amplitude of the fluctuations, $\sigma/\mu$, is higher for larger $\Delta$ values for all densities. This is the consequence of mean values of $I_{sat}$ falling faster than the fluctuation levels. Also, $\sigma/\mu$ values increase roughly by $25\%$ over the transition at all radial positions. In Fig. \ref{fig:raw_data} (b), the autocorrelation time, $t_{AC}$, of the signal is displayed, indicating the characteristic time scale of fluctuating structures in the data set. In the near SOL, a very clear effect of increasing density can be seen: $t_{AC}$ becomes longer by over a factor of five when the highest densities are achieved. Instead, the effect in the far SOL seems to be moderate, although no measurements for $\bar{n}_e > 3 \cdot 10^{19}$ m$^{-3}$ are available in the far SOL. As a result, the radial profile changes: below the transition, the duration of filaments increases with the distance to the separatrix. After the transition, the longest lasting filaments are found very close to the separatrix, and $t_{AC}$ decreases to previous values as $\Delta$ is increased. In Fig. \ref{fig:raw_data} (c), the skewness, S, of $I_{sat}$ is represented. $S$ is the third statistical moment of the data set, indicating the symmetry of the probability distribution function (PDF). Positive values of $S$ are associated to the existence of positive tails in the PDF of the signal, which are indicative of a predominance of positive fluctuations (blobs) over negative ones (holes). In the plot, positive values of $S$ are found in the whole parameter space, which is a universal feature in the SOL of most fusion devices \cite{Dippolito10} and measured values are similar to those previously reported in AUG \cite{Nold10} for equivalent densities: Increasing with distance to the separatrix in the $S\in[0.5,2]$ range. As the density is increased over the lowest values ($\bar{n}_e > 1.5 \cdot 10^{19}$ m$^{-3}$), $S$ values are also increased over the whole SOL and specially for the highest values of $\Delta$. However, further increases in $\bar{n}_e$ - including the transition - do not seem to have a significant effect on the skewness. In Fig. \ref{fig:raw_data} (d), the blob detection rate, $\nu_b$, defined as the ratio between the number of blobs detected and the duration of the data sample, is displayed. As can be seen, the behavior of $\nu_b$ almost mirrors that of $t_{AC}$: for low densities, high $\nu_b$ values are found close to the separatrix, and then decay with $\Delta$. As $\bar{n}_e$  increases, blobs become less frequent at the lowest $\Delta$ region, but seem mostly unaffected in the far SOL. The fraction of the sample corresponding to ``blob time'', which can be calculated as $f_{blob} = \nu_b t_{AC}$ is also affected by the transition: under $\bar{n}_\textnormal{HDT}$, it shows values in the range of $f_{blob} = 2-3 \%$ while over $\bar{n}_\textnormal{HDT}$ it increases to $f_{blob} = 4-7 \%$. Incidentally, these values are always sufficiently low to consider the evolution of $\nu_b$ real and not just the trivial result of less blobs fitting in a limited time interval. Summarizing, Figs. \ref{fig:raw_data} (a), (b) and (d) show clearly different properties above and below the same $\bar{n}_\textnormal{HDT} \simeq 2.5 \cdot 10^{19}$ m$^{-3}$ edge density value which determines the transition between SOL radial profiles shown in Fig. \ref{fig:figprofiles}.\\

\begin{figure}
	\centering
		\includegraphics[width=\linewidth]{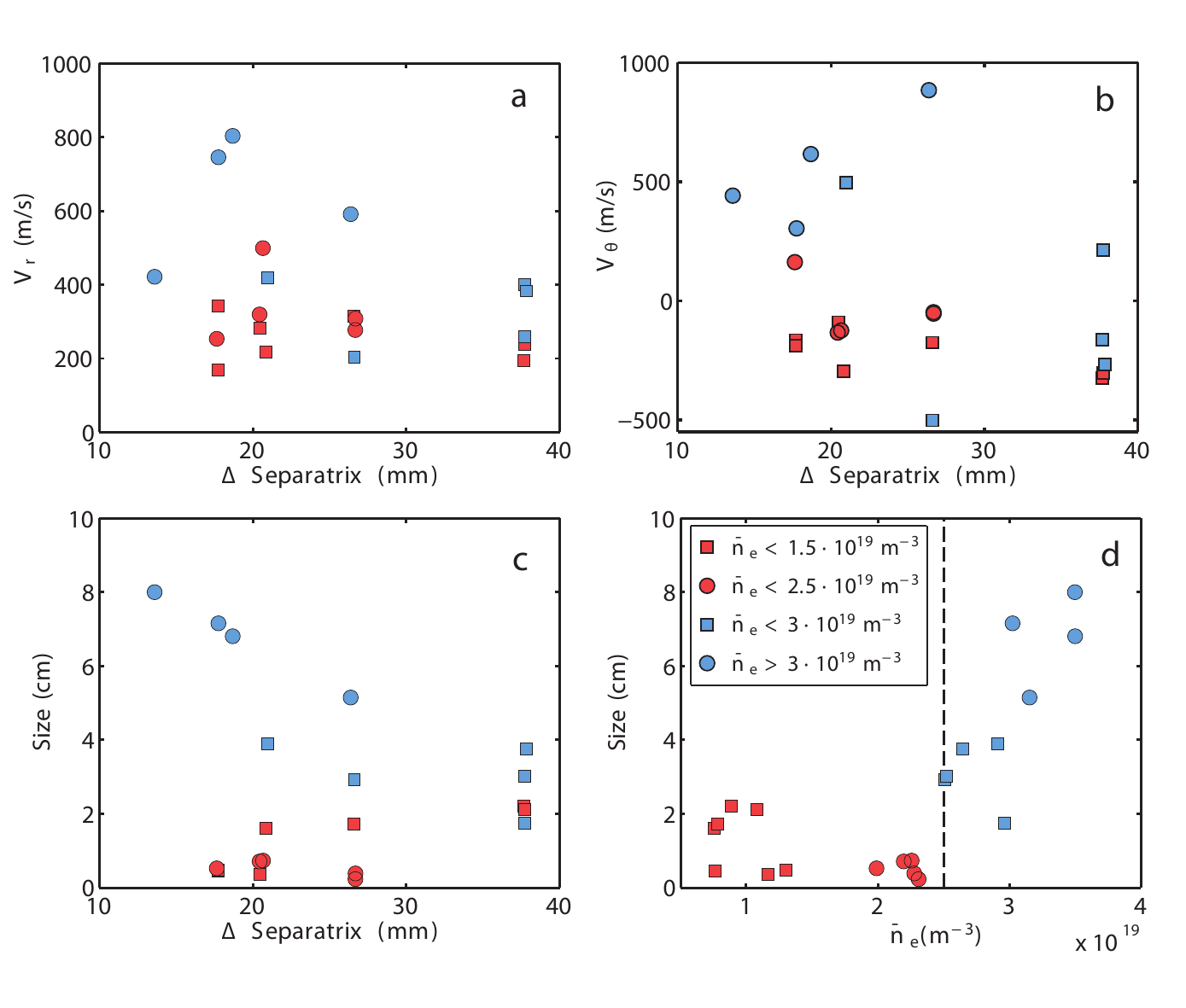}
	\caption{\textit{Results of the conditional analysis. Points are classified in the same four groups of increasing densities defined in Fig. \ref{fig:raw_data}. The direction of radial and binormal velocities directions is defined as in Fig. \ref{fig:diag} (\textit{i.e., $v_\theta<0$ is directed towards the LFS target}). The dashed line in Fig. (d) indicates the $\bar{n}_e = \bar{n}_\textnormal{HDT}$ threshold for the transition.}}
	\label{fig:fig2}
\end{figure}

Once the basic properties of the turbulence have been described, the propagation of filamentary structures is analyzed as explained in section \ref{anatech}. The results of this analysis are displayed in Fig. \ref{fig:fig2}: As can be seen in Fig. \ref{fig:fig2} (a), the radial velocity of filaments, $v_r$, rises with density, especially at the innermost positions of the SOL, where the values are roughly increased by a factor of $4$ after the transition. The binormal velocity of filaments, $v_\theta$, is shown in Fig. \ref{fig:fig2} (b). After the density transition, a strong change can be observed in the profile of $v_\theta$: at low densities $v_\theta$ is negative (i.e., goes in the ion diamagnetic direction). This result is consistent with previous probe and Doppler reflectometry measurements carried out in the same $\Delta$ region in a similar discharge with $f_{GW} \simeq 0.42$ \cite{Nold10, Conway06}. Instead, for high densities, $v_\theta$ takes large positive values for $\Delta < 30$ mm and then displays a strong velocity shear layer after which $v_\theta$ becomes negative again. Finally, in Fig. \ref{fig:fig2} (c) and (d), the evolution of filament size, $\delta_b$, is displayed. In Fig. \ref{fig:fig2} (c), the change of the radial structure across the transition is evident: as the $\bar{n}_e =\bar{n}_\textnormal{HDT}$ threshold is surpassed, filaments close to the separatrix become progressively larger,their size increasing up to an order of magnitude with respect to pre-transition values. As in Fig. \ref{fig:raw_data} (b) the radial trend is reversed, and $\delta_b$ decreases with $\Delta$, becoming closer to pre-transition values for $\Delta \simeq 40$ mm. The effect of the transition is seen even more clearly in Fig. \ref{fig:fig2} (d), where $\delta_b$ is represented as a function of $\bar{n}_e$. After the transition $\delta_b$ starts to increase roughly linearly with  $\bar{n}_e$. This increase of almost an order of magnitude results from the simultaneous rise in flight times (which could already be seen in the raw data in Fig. \ref{fig:raw_data}) and propagation speeds.

\subsection{Relation to divertor detachment}

\begin{figure}
	\centering
		\includegraphics[width=.4\linewidth]{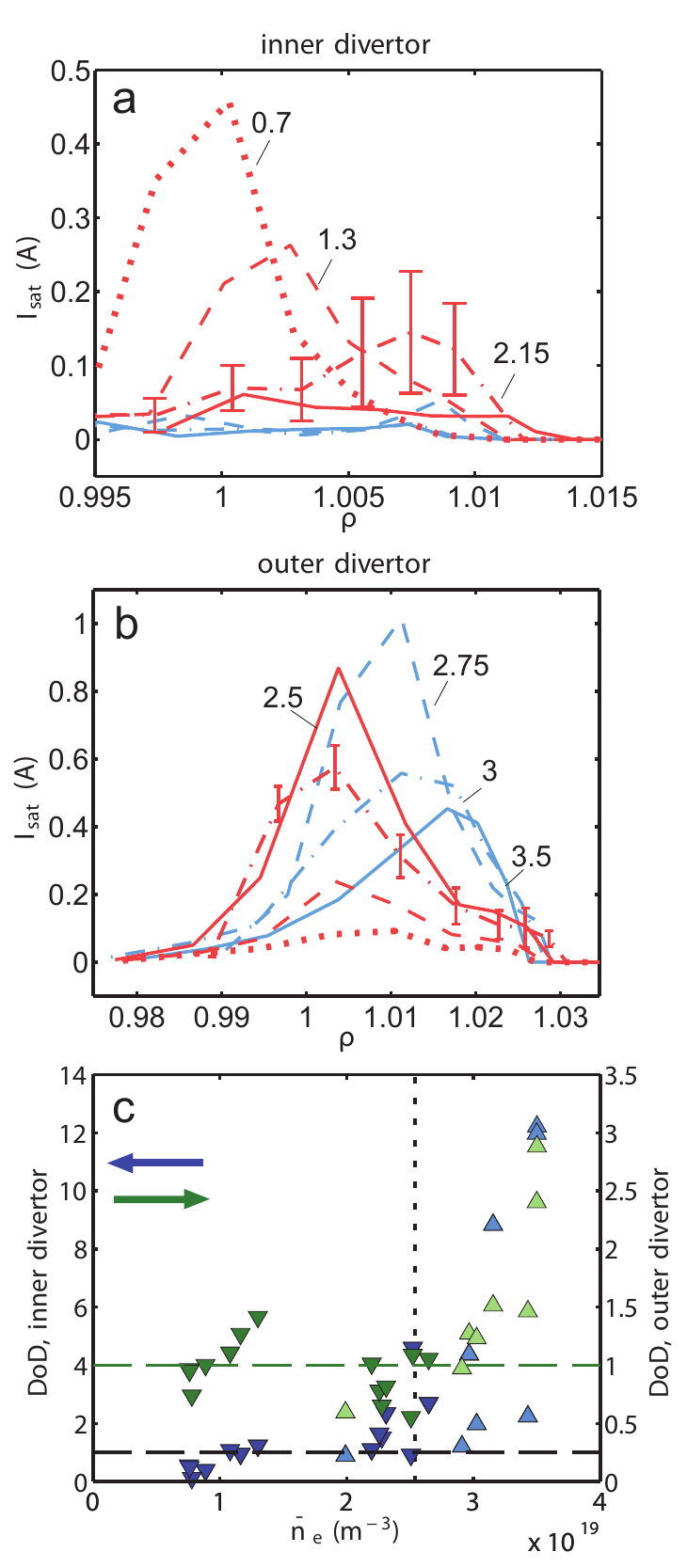}
	\caption{\textit{Divertor evolution. (a) and (b) show the radial profiles of $I_{sat}$ as measured by the Langmuir probes in the divertor tiles for the same values of $\bar{n}_e$ used in section \ref{prof} to discuss the midplane profile evolution. Color code as in Fig. \ref{fig:figprofiles}. For convenience, the corresponding densities (in $10^{19}$ m$^{-3}$) are displayed next to the curves. In (c), the degree of detachment of both divertors is represented as a function of $\bar{n}_e$. Blue/green markers stand for inner/outer divertor. Light/dark colors indicate positive/negative $v_\theta$. As indicated by the direction of the triangle, $v_\theta > 0$ means upwards (electron diamagnetic direction) in the LFS midplane. Dashed lines at each axis indicate the approximate value for which detachment starts. The vertical dashed line indicates $\bar{n}_\textnormal{HDT}$.}}\label{fig:div}
\end{figure}

When analyzing the evolution of the parallel transport of the SOL over the transition in section \ref{prof}, indications were found that the divertor begins to detach for the highest values of $\bar{n}_e$. In order to clarify the relation between the density transition and divertor detachment, a more systematic study is carried out with the two arrays of divertor probes: In Figs. \ref{fig:div} (a) and (b), the $I_{sat}$ profiles measured by these probes are displayed for both divertors at the same times when the midplane profiles of Fig. \ref{fig:figprofiles} were obtained. The same color code is used to indicate $\bar{n}_e$ values. Again, red and blue curves correspond to below and above the transition respectively. As can be seen, the outer divertor displays a clear transition from the sheath limited to the detached regime: at the lowest values of $\bar{n}_e$, the profile peaks close to the separatrix, indicating an attached divertor, and $I_{sat}$ at the target changes weakly with $\bar{n}_e$. Then, between $\bar{n}_e = 1.5 \cdot 10^{19}$ m$^{-3}$ and $\bar{n}_e = \bar{n}_\textnormal{HDT}$, there is a fast increase of $I_{sat}$ at the target (over a factor of five), indicating the onset of the high recycling phase. Finally, for $\bar{n}_e > \bar{n}_\textnormal{HDT}$  $I_{sat}$ on the target begins to decrease, resulting in a less peaked profile and indicating the beginning of divertor detachment close to the separatrix. The inner divertor undergoes the same transition at lower densities and only shows a well attached regime for the lowest $\bar{n}_e$ value. Above $\bar{n}_e \simeq 1 \cdot 10^{19}$ m$^{-3}$, the detachment phase starts, $I_{sat}$ on the target is reduced and the profile becomes broader while the peak shifts outward. Finally, the $\bar{n}_e \geq \bar{n}_\textnormal{HDT}$ transition seems to happen around the time when the inner divertor completely detaches and the particle flow to the divertor is practically reduced to zero. Further increases have no effect on the profile (blue curves are merged together at marginal values of $I_{sat}$). These observations are in good agreement with previous studies of the onset of divertor detachment in AUG \cite{Potzel14}.\\

In order to provide a more quantitative description of the divertor evolution with $\bar{n}_e$, the Degree of Detachment (DoD) is calculated during each MEM plunge. The DoD is the ratio between the measured flux to the divertor and the prediction of the 2-point model \cite{Potzel14}. Therefore, for an attached divertor DoD$ = 1$ and after the onset of detachment DoD$ \gg 1$. For this, the $I_{sat}$ measurements of the fixed divertor Langmuir probes are summed up separately for the inner and outer divertor targets. The equal spacing of the probes along the target guarantees that the sum is proportional to the integrated flux, given the $I_{sat}$ profile is well resolved by the Langmuir probes. The sum is normalized by $\bar{n}_e^2$ to account for the two point model \cite{Stangeby} scaling with upstream density. Finally, the DoD is derived by taking the ratio of the normalized sum at a given time point with the normalized sum at a well attached reference time. No changes in heating power/power entering the SOL are taken into account. The resulting values are	 represented in Fig. \ref{fig:div} (c). As can be seen, the results are in good agreement with the profile evolution: in the inner divertor, detachments sets in for $\bar{n}_e > 2.2 \cdot 10^{19}$ m$^{-3}$ , while in the outer divertor DoD $> 1$ is only achieved at $\bar{n}_e \simeq 3 \cdot 10^{19}$ m$^{-3}$. Also, the DoD values in the inner divertor are much higher (note the different scale in the figure), indicative of the more complete detachment observed there. The colored triangles indicate the binormal direction of the filament propagation, as measured by the MEM probe array. Fig. \ref{fig:div} shows a strong coincidence of the density range in which filaments and profiles change, and that for which the detachment of the LFS divertor starts. This result confirms with a much greater level of detail the early observations of AUG divertor \cite{Rohde96}. Besides, the change in sign of the binormal filament propagation is obviously taking place when the DoD of the outer divertor becomes greater than one. This novel experimental result will be addressed in the discussion.\\

\subsection{Changes in transport}\label{transport}

\begin{figure}[t!]
	\centering
		\includegraphics[width=\linewidth]{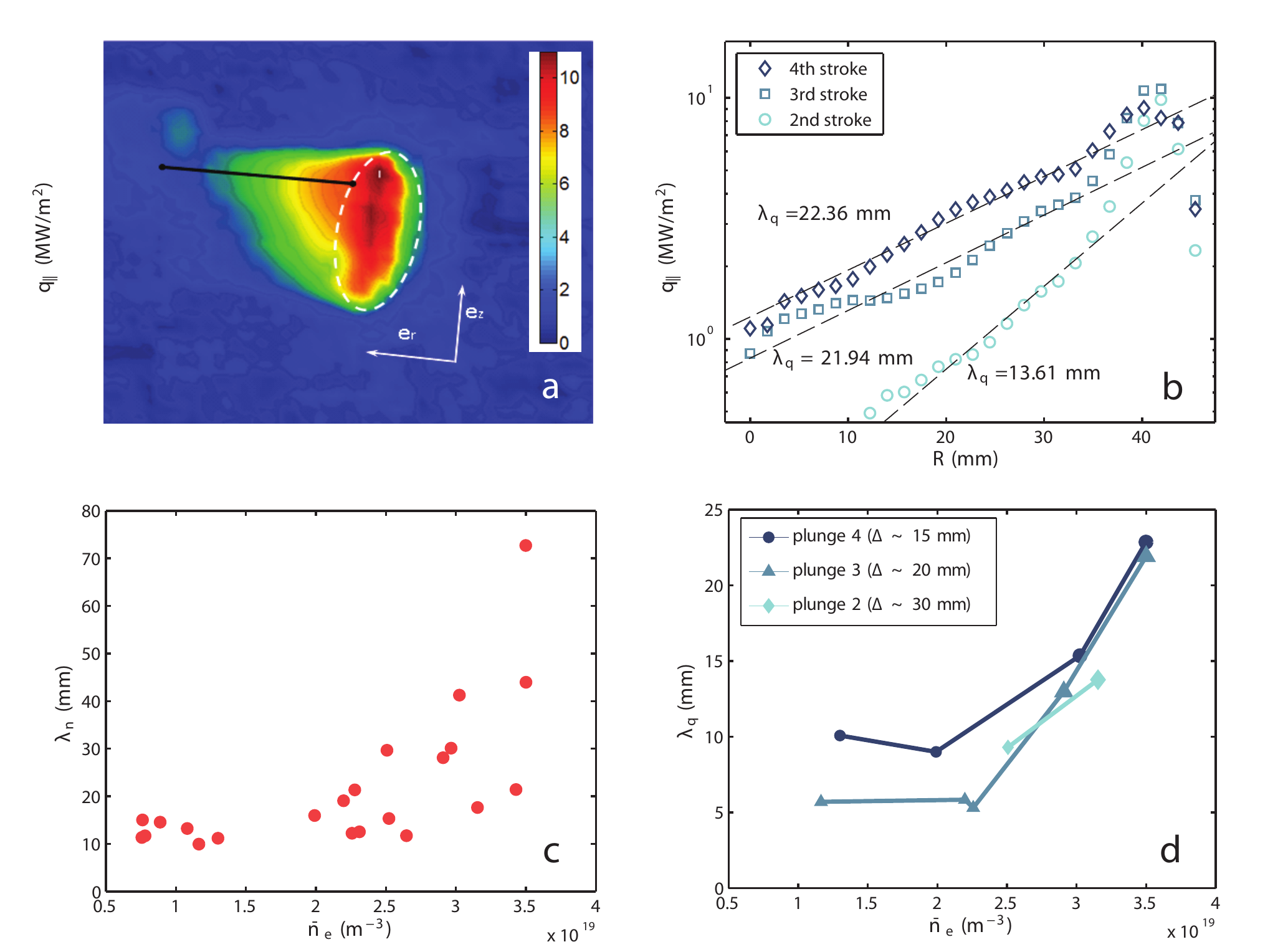}
	\caption{\textit{Changes in SOL transport. In (a), a typical 2D plot of $q_{||}$ on the MEM probe head surface is displayed, as calculated from the IR camera data. The  contour of the last cylindrical section of the probe is indicated as a white dashed line. The two axis represent local vertical and radial directions. The radial sampling region is displayed as a black line. In (b), radial profiles of $q_{||}$ are displayed for the three last plunges of discharge \#29887. Corresponding $\lambda_q$ values are fitted and displayed. In (c), $\lambda_n$ values calculated from LiB data in the $\rho \in [1,1.03]$ range from Fig. \ref{fig:figprofiles}, as a function of $\bar{n}_e$. In (d), $\lambda_q$ values calculated from the IR $q_{||}$, as a function of $\bar{n}_e$.}}
	\label{fig:transport}
\end{figure}

Once the change of filaments has been characterized across the density threshold, the question about its relevance for SOL transport remains: has the change on filament structure an impact on the heat and particles being transferred to the first wall? The changes in the profile observed in LiB data (displayed in Fig. \ref{fig:figprofiles}) indicate that the SOL broadens significantly after the transition, which is a strong indicator of an increased perpendicular transport. In order to describe this broadening in a more quantitative manner, the density e-folding length, $\lambda_n$, is calculated from LiB data. For this, an exponential decay is fitted in the $\rho \in [1,1.03]$ range of the profile, roughly corresponding to the region between the separatrix and the limiter shadow. The results, displayed in Fig. \ref{fig:transport}(c), confirm the previous finding from Fig. \ref{fig:fig2}(d): $\lambda_n$ remains constant for $\bar{n}_e < \bar{n}_\textnormal{HDT}$ and then rises with $\bar{n}_e$, becoming about five times larger than those found at lower densities.\\

A similar analysis can be carried out on thermal transport using the data from the IR camera which overlooks the MEM. As explained in section \ref{Exp}, this diagnostic delivers $q_{||}$ onto the surface of the probe head, allowing the fitting of exponential curves to the radial profiles of $q_{||}$ , thus yielding the parallel heat flux e-folding length, $\lambda_{q}$, on the outer midplane. This analysis is presented in Fig. \ref{fig:transport}: In Fig. \ref{fig:transport} (a), a typical 2D plot of IR measurements is displayed together with the radial and vertical directions and the elliptical projection of the last cylindrical section of the probe head (indicated in Fig. \ref{fig:diag}). This section, which can be easily fitted in the IR silhouette of the probe head, is used to obtain a local spatial reference, both for determining the size of the measured features and  to define the local radial and vertical directions. In order to have a consistent measurement of the radial heat transport in the SOL, the calculated $q_{||}$ is sampled along a radial line, indicated as a black segment in the plot. The probe head front features a much more complex geometry due to the several pin terraces, and can not be used for the determination of $\lambda_q$. As an example, the three last plunges of discharge \#29887 are shown in Fig. \ref{fig:transport} (b): in it, the radial decay of $q_{||}$ along the cylindrical body of the probe head is roughly exponential, and a clear value of $\lambda_q$ can be deduced from these measurements. The results of the analysis are presented in Fig. \ref{fig:transport} (d), where the $\lambda_{q}$ is displayed as a function of $\bar{n}_e$. This analysis can not be performed for the lower density values ($\bar{n}_e < 2 \cdot 10^{19}$ m$^{-3}$), for which the raw data is too low in the cylindrical region to obtain reliable $\lambda_q$ values. As a result, only data points featuring sufficient density are displayed including third and fourth strokes, as well as the second stroke measurements from the highest density discharges. As seen in the plot, $\lambda_q$ displays the same general behavior as $\lambda_n$, including a sharp increase after $\bar{n}_e = \bar{n}_\textnormal{HDT}$. In this case, the e-folding length is increased by a factor of $2-3$ across the transition. For low density, values from the third plunge are half of those from the fourth (corresponding to $\Delta \simeq 20$ mm and $\Delta \simeq 15$ mm, respectively). After the transition, they  become equal indicating a broadening of the SOL in that radial range.\\

In conclusion, the SOL broadens clearly across the transition, and particle and power load profiles become flatter by a factor of 5 and 3, respectively. Also, after the transition $\lambda_n$ seems to become larger than $\lambda_q$ at least by a factor of 2. This contrasts with measurements carried out for type-I ELMs at equivalent densities ($f_{GW} \simeq 0.6$) \cite{Herrmann07}, in which $\lambda_q$ and $\lambda_n$ show similar values.\\

\section{Discussion}\label{Discussion}

The results presented in the previous section have clearly demonstrated the existence of a transition in SOL transport at high densities. Indeed, as can be observed in Fig. \ref{fig:figprofiles}, for edge densities over $\bar{n}_\textnormal{HDT} \simeq 2.5 \cdot 10^{19}$ m$^{-3}$ the profile structure transits from a two-region SOL (dominated by parallel transport in the vicinity of the separatrix and by perpendicular transport in the far SOL) to a single structure in which the slope remains roughly constant across the whole region. As a result, the density rises in the whole SOL indicating an increased importance of the perpendicular particle transport with respect to the parallel one. This process is equivalent to those already found in the literature \cite{Labombard01, Garcia07}, where the same transition is reported. The value of the density threshold $\bar{n}_\textnormal{HDT}$ corresponds to a Greenwald fraction of $f_{GW} \simeq 0.45$, also similar to those from other machines (see introduction). In literature \cite{Myra06}, the transition is attributed to a change in the filament regime triggered by the increase of $\Lambda$. As a result, convection - previously only dominant in the far SOL - would become the dominant transport mechanism in the whole SOL, thus removing the fast conductive parallel transport observed in the near SOL in the low density cases. Nevertheless, the observed changes in the profiles are not sufficient to verify this hypothesis: To do so, it is required to prove that a) filaments undergo a regime change at the transition point, b) such transition leads to a clear increase in filamentary transport and c) this is related to an increase of $\Lambda$.\\

\subsection{Evidence of a density transition on filaments}

The first point has been sufficiently documented in the previous section: both the general statistical properties of turbulence and the conditional analysis of filaments reveal a clear change in these structures at densities higher than $\bar{n}_\textnormal{HDT}$. These changes are particularly pronounced in the region of observation closest to the separatrix ($\Delta \in[10,25]$ mm), corresponding to a flux coordinate of $\rho \in [1.01,1.03]$. As can be seen in Fig. \ref{fig:figprofiles}, this means that the region where the main profile change takes place is only marginally probed during the deepest plunges. Instead, the observed structures are mostly the filaments which propagate ballistically through the convection region. Still, since both the autocorrelation time, $t_{AC}$, and size, $\delta_b$, of the blobs seem to increase towards the separatrix, it is likely that a similar if not more pronounced process takes place between $\rho\in[1.005,1.01]$.  Also, as displayed in Fig. \ref{fig:fig2} (a) and (b), radial filament propagation becomes faster and a velocity shear layer, where the propagation direction reverses appears in the binormal direction. The increase in the size displayed in Fig. \ref{fig:fig2} (c) and (d) must be interpreted as the characteristic perpendicular size: as explained in the Appendix, no information can be obtained in the dimension of the blob perpendicular to $\bf{V}_\bot$. In this context, the fact that $\delta_b$ values of the order of and even larger than the SOL width is difficult to interpret, and could be the result of an over estimation of $\bf{V}_\bot$. One alternative explanation would be that filaments become elongated in the direction of propagation, effectively resulting in a SOL dominated by streamers. This would be in good agreement with the simulations of the combined effect of many blobs in transport carried out by Myra et al. \cite{Myra07}. They found that for high densities, the SOL could be dominated by radially elongated convective structures, which could explain the observed sizes. Since not much experimental data are available for this high density L-mode, new experiments with 2D diagnostics (such as GPI) will have to be carried out in order to sort this out. Nevertheless, it can be stated that the results are at least qualitatively valid, as both $v_r$ and $\delta_b$ fall well within the typical range of values measured in many other machines: in order to compare to the multi-machine review presented by D'Ippolito et al. \cite{Dippolito10}, the ``stable blob scale'' normalization \cite{Krash07} is used, $\hat{\delta} = \delta_b/\delta_*$ and $\hat{v} = v_r/v_*$, where\\

\begin{equation}
\delta_*=\rho_s^{4/5}L_\parallel^{2/5}/R^{1/5},  \quad  v_*= c_s (\delta_*/R)^{1/2}
\end{equation}

and $\rho_s$ stands for the ion gyroradius. By doing so, typical values of $\hat{\delta} \simeq 2$, $\hat{v} \simeq 0.1$ and  $\hat{\delta} \simeq 8$, $\hat{v} \simeq 0.5$ are obtained below and over the transition, respectively. As can be seen in Fig 27 in \cite{Dippolito10}, these values are clearly in the same range than those reported elsewhere. \\

\subsection{Effects on transport}

\begin{figure}
	\centering
		\includegraphics[width=.8\linewidth]{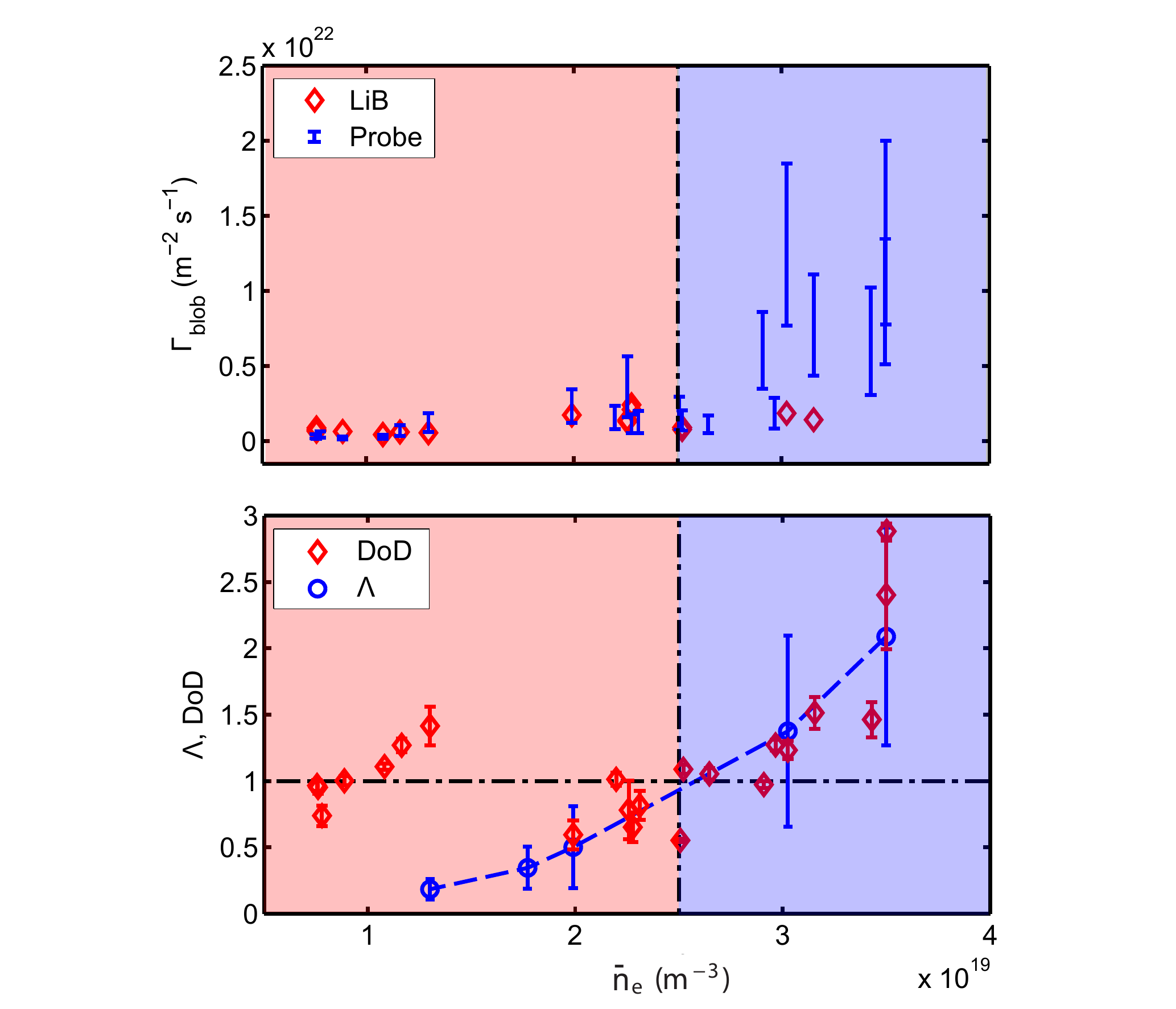}
	\caption{\textit{Summary of results: Top, blob related particle transport $\Gamma_{blob}$ is displayed as a function of edge line density. Blue lines represent the upper and lower bounds for $\Gamma_{blob}$ as measured by the MEM. Red diamonds represent $\Gamma_{blob}$ as measured by Lithium beam spectroscopy (LiB). Bottom, the degree of detachment (DoD) of the outer divertor and the outer midplane parallel normalized collisionality, $\Lambda$, are represented respectively as red diamonds and blue circles.}}
	\label{fig:resumen}
\end{figure}

In section \ref{transport} it has  already been shown how the transition leads to a clear broadening of the SOL, indicating enhanced levels of perpendicular transport. In order to determine the role of filaments in this enhancement, an estimation of the convective transport associated to blobs has been carried out. This is defined as $\Gamma_{blob}=\left\langle n_{blob}\right\rangle \left\langle v_{r,blob}\right\rangle$, where $\left\langle n_{blob}\right\rangle$ and $\left\langle v_{r,blob}\right\rangle$ are respectively the conditionally averaged density and radial velocity of blobs, calculated as explained in section \ref{anatech}. The evolution of $\Gamma_{blob}$ with $\bar{n}_e$ is shown in Fig. \ref{fig:resumen} (top), both for LiB and MEM data. Interestingly, the good agreement found in Fig. \ref{fig:profiles_multi} can be also found here for densities below $\bar{n}_\textnormal{HDT}$, where the LiB diagnostic can be considered reliable for the measurement of fluctuations. As can be seen, filamentary transport is increased by almost an order of magnitude over the transition, up to values in the order of $\Gamma_{blob} \simeq 10^{22}$ m$^{-2}$ s$^{-1}$: the filaments do not only increase in amplitude, but also enhance the perpendicular transport. This sharp increase in convective transport is consistent with the broadening of the profiles, but it still needs to be proven that the contribution of $\Gamma_{blob}$ is enough to change the whole transport in the SOL. In this sense, it must be pointed out that the values displayed in Fig. \ref{fig:resumen} (top) only represent $\Gamma_\bot$ during blob time. In order to get a more realistic estimation of the average blob transport, $\Gamma_{blob}$ must be factored by the ``blob time'' fraction, $f_{blob}$. This yields values in the order of $\Gamma_{blob}f_{blob} \simeq 10^{21}$ m$^{-2}$s$^{-1}$ after the transition. This value represents at least a substantial fraction of the perpendicular flow calculated by SOLPS simulations \cite{Leena12} in the attached regime for similar radial positions (of $\Gamma_\bot \in [0.75, 2.5] 10^{21}$ m$^{-2}$s$^{-1}$), which indicates qualitatively the relevance of filamentary transport. However, this is just a crude estimation, and a more detailed analysis, including a comparison of $\Gamma_{blob}$ with other particle fluxes in the SOL, such as $\Gamma_\parallel$, the contribution of diffusion or the ionization source term \cite{Lipschultz05}, or a study in which the consequences of this convection in heat transport are compared with calorimetry measurements \cite{Labombard01} are still required in order to state that filaments are responsible for the changes in the profiles. This task will be addressed in future work.\\


\subsection{Comparison to analytic model}

Finally, the validity of the filament transition model proposed by Myra et al. \cite{Myra06} is considered. To do so, the values of normalized collisionality $\Lambda$ are calculated following the definition of equation (\ref{eq1}). Since the filament circuit is closed along the field lines, the relevant $\Lambda$ values are those associated to the fluctuating values of density and temperature. Therefore, $T_e \simeq 15$ eV and $T_i \simeq 90$ eV are assumed and an average value of blob densities is obtained from LiB data. The connection length, $L_c$, to the outer divertor is used. The results are displayed in Fig. \ref{fig:resumen}(bottom): $\Lambda$ values, covering a range of around an order of magnitude, cross the $\Lambda = 1$ line at $\bar{n}_\textnormal{HDT}$, very much like the other transitions in profiles and filament shape. As explained in the introduction, at $\Lambda = 1$, the characteristic time of parallel transport becomes larger than the mean free path of ion-electron collisions, disconnecting the midplane from the divertor and triggering the transition. This result indicates that a mechanism similar to the convection feedback loop \cite{DIppolito06} could be responsible for the transition. It must be pointed out that this explanation doesn't take into account the role of main chamber wall recombination and ionization of neutrals, which also has been proposed \cite{Neuhauser02, Lipschultz05} as an alternative explanation for this process. Both views are complementary, but a detailed discussion on their relative merits is out of the scope of this work and will be left for a future paper.\\

Despite the general good agreement between the described phenomena and the predictions of the model, there are still a number of caveats with these results: In the first place, no measurements are available for the fluctuating values of $T_i$ and $T_e$. Although this is probably an acceptable approach for the calculation of $n_e$ from $I_{sat}$ (since $c_s \propto (T_e+T_i)^{1/2}$), moderate fluctuations of $T_e$ may be relevant for $\Lambda$ since $\nu_{ei} \propto n_e \cdot T_e^{-3/2}$. On the other hand, the absolute calculation of $n_e$ also involves some uncertainty since the real collection area of the probe is not known and the assumed constant $A_{eff} \simeq 5$ mm$^2$ does not take into account the effects of sheath expansion. By comparison of a three and four parameter fitting \cite{Gunn96} of the I-V curves measured in similar conditions, the error induced by this last effect is estimated in the range of $10-30\%$. More importantly, the estimations of $T_e$, $T_i$ and $n_e$ values are only valid for the midplane, while strong changes in collisionality are to be expected between the X point and the divertor as $\bar{n}_e$ is increased. In particular, the detachment of the divertors must have a determinant influence on the connection of the filaments: not only the two point SOL approach is no longer valid, as seen in the deviation of the density profiles at the X-point from those at the midplane (Fig. \ref{fig:profiles_multi}), but also the presence of a cold region in front of the wall may interrupt the filament circuit, thus forcing the end of the sheath connected regime regardless of the values of $\Lambda$. As can be seen in Fig. \ref{fig:div}, the beginning of the divertor detachment affects the $\rho \in [1,1.03]$ region, which roughly corresponds to $\Delta < 30$ where the main changes for filaments take place. Also, in Fig. \ref{fig:resumen} (bottom), the evolution of the DoD on the outer divertor displays a very similar evolution indicating that both the beginning of the detachment and the collisionality transition happen roughly at the same time.\\

\subsection{Relation to divertor detachment}

In this sense, the discovered relation between the filament transition and the detachment of the outer divertor is very relevant and raises an intriguing question about the direction of causality in the observed events: one possible mechanism is that the increase of $\Gamma_{blob}$ and density in the SOL leads to a rise in the global collisionality $\Lambda$ along field lines, disconnecting the filaments from the divertor and leading to an enhanced convective radial transport. This would spread the power loads arriving into the divertor and cause its detachment or at least speed up the transition from partial to complete detachment along the entire target creating an enhanced power loss channel. However, the process could also work in the opposite direction: the beginning of the divertor detachment, caused by reasons unrelated to filamentary transport in the SOL, creates a low temperature region in front of the targets which would greatly increase collisionality and thus disconnect the filaments regardless of the $\Lambda$ values in the midplane. This effect would start close to the separatrix (where the detachment starts) and propagate then towards greater $\rho$ values. As a result, filaments in the midplane would transit to the collisional regime and increase the radial transport in the observed way. Both possibilities seem realistic and it is not possible to determine the direction of causality with current data. So far, the discussion about the effect of detachment has been focused on the outer divertor. Since it has the lowest $L_c$ to the midplane, and given the clear connection between its DoD and the formation of the velocity shear it would be expected to have greater influence. However, the question remains whether the key factor in the filament behavior is the beginning of the detachment of the outer divertor or the achievement of the total detachment in the inner one, since both happen roughly at the same times.\\

Another midplane effect of divertor detachment could be the emergence of the shear layer after the transition. A possible explanation for this would be, again, the electrical disconnection from the divertor: for low densities, the attached condition creates a $\nabla_r T_e$ which, due to the sheath potential drop, leads to a $E_r > 0$ in front of the target plates. In this low density condition, parallel gradients are very low and $E_r$ at the midplane is strongly influenced by the divertor temperature \cite{Stangeby}. Therefore, an $E\times B$ drift appears in the ion diamagnetic direction, as observed experimentally. Over $\bar{n}_\textnormal{HDT}$, the high recycling regime causes $T_e$ to drop to a few eV for $\rho < 1.03$. As a result, the $E_r$ in front of the target is reduced and, more importantly, the $E_r$ at the midplane is disconnected from the divertor. That radial range corresponds to the region where inversion of $v_\theta$ is observed. In good agreement with that, $\phi_p$ profiles in the X-point are flattened after the transition. Instead, for $\rho > 1.03$ the divertor is still not detached, $T_e$ remains relatively unaffected at the divertor and $v_\theta$ doesn't change sign in the midplane.\\


\section{Conclusions}\label{conclusion}

The first conclusion of this work is that the high density transition in AUG displays the same behavior as reported from other machines in the reviewed literature: SOL density and $q_\parallel$ profiles broaden at densities above a threshold value of $\bar{n}_\textnormal{HDT} = 2.5 \cdot 10^{19}$ m$^{-3}$, corresponding to $f_{GW} \simeq 0.45$. This change is observed with several diagnostics: The broadening of $n_e$ profiles is measured with LiB and Langmuir probes in the midplane and X-point reciprocators, while the broadening of the $q_{||}$ profile is measured by the IR camera on the MEM probe head. The agreement between the different diagnostics is remarkable. In particular, LiB data show how the structure of density profiles transits from a dual shape with a steep gradient close to the separatrix to a single region one, in which the whole SOL shows the gradient observed in the far SOL at lower densities.\\

However, due to the combined use of SOL diagnostics, a number of additional findings can be reported: In the first place, clear evidence has been provided that such variations in the profiles are accompanied by changes in the structure of filaments in the outer midplane. As the density surpasses the transition threshold, filaments become larger and denser, increase their radial propagation speed and their detection frequency is reduced. This transition coincides with an increase in radial transport. This was measured both on the changes of $\lambda_n$ and $\lambda_{q}$, which rise respectively by a factor of 5 and 3, and on the average  filamentary transport $\Gamma_{blob}$ measured by the multipin probe head of the MEM, which is increased by more than an order of magnitude leading to effective particle fluxes up to $\Gamma_{blob}f_{blob} \simeq 10^{21}$ m$^{-2}$s$^{-1}$. Finally, all these changes coincide with the detachment of the outer divertor, as indicated by the degree of detachment and radial ion saturation profiles measured by the flushed mounted Langmuir probe arrays on the targets. The outer divertor is the one with the shortest connection length to the midplane and the last to detach. Therefore, it can be expected to be the main electrical interface between filaments and the solid wall. Probably as a result of this, the poloidal rotation of filaments changes from the ion to the electron diamagnetic drift direction.\\

In order to explain the underlying physical mechanism, the validity of some models proposed in the literature has been investigated \cite{Myra06,Garcia06,DIppolito06}. According to these models, these changes would be the result of a transition in the filament regime, governed by collisionality along the filament, $\Lambda$. An estimation of the values of $\Lambda$ showed it increases with density, $\bar{n}_e$, and crosses the critical $\Lambda = 1$ value when $\bar{n}_\textnormal{HDT} = 2.5 \cdot 10^{19}$ m$^{-3}$ is reached. This result is consistent with the model proposed by Myra et al. \cite{Myra06}. However, these values only correspond to estimations on the midplane and must thus be taken with caution. A more meaningful calculation of $\Lambda$ would require to take into account the variations of $n_e$, $T_i$ and $T_e$ along the whole field line (including the divertor) as well as the effect of neutrals and detachment. Such quantitative calculations are considered beyond the scope of this work and left for future studies. Nevertheless, given the qualitative agreement between the experimental results and the proposed models, an increase in convective transport can be considered as the mechanism governing the transition. The direction of causality between midplane profiles and divertor detachment (i.e., whether increased filamentary transport triggers the detachment or increased resistivity due to divertor detachment triggers the filament transition) will be addressed in future work.\\

Summarizing this work, a consistent picture is provided in which many experimental findings are brought together under a coherent description linking filament models, midplane perpendicular/parallel transport balance and divertor detachment. These findings may turn out of great practical importance for the design and operation of next generation tokamaks: since these devices will operate at high densities and with detached divertors, transport in the SOL can be expected to be dominated by filaments in the high density regime described in this work. Therefore, the extrapolation of SOL transport based on experimental observations conducted in present tokamaks with attached divertors and low collisionalities may be seriously underestimating the future particle and heat fluxes onto the main chamber wall. In this sense, it is particularly important to determine whether the key parameter governing the transition is the $\Lambda$ parameter at the main chamber or at the detached divertors, since nominal regimes in ITER will feature partially detached divertors but still a very hot SOL in the midplane (with a relatively low $\Lambda$). These regimes can't be achieved simultaneously in present day machines, excluding the possibility of direct extrapolation. Therefore, ITER operation will require a better understanding of the whole process, including a working conceptual model with some degree of predictive capability.\\


\section*{Acknowledgements}

The authors would like to thank S. Potzel for the useful discussions. This work was supported by EURATOM and carried out within the framework of the European Fusion Development Agreement. The views and opinions expressed herein do not necessarily reflect those of the European Commission. \\

\section*{Appendix: Filament Propagation Analysis}\label{app}

Blob velocity and size calculations based on conventional blob analysis techniques using the correlation between pins yielded non-physical results, including irrealistically large velocities (and sizes). Considering the initial results, featuring filaments with perpendicular sizes in the range of several centimeters, an obvious source of error was the fact that filaments are typically larger than the distance between pins (in the range of a few mm). Therefore, some model of the simultaneous interaction of the blob with the whole pin array is required in order to correctly interpret probe data. Such a simple model was then introduced, based on the following assumptions: 

\begin{itemize}
\item [-] The density distribution of the blob can be roughly approximated as an ellipse.
\item [-] The size of the blob is larger than the projection of the distance between the pins on the axis perpendicular to the propagation.
\item [-] The blob is symmetric with respect to axis parallel to the direction of propagation.
\end{itemize}

\begin{figure}[t!]
	\centering
		\includegraphics[width=\linewidth]{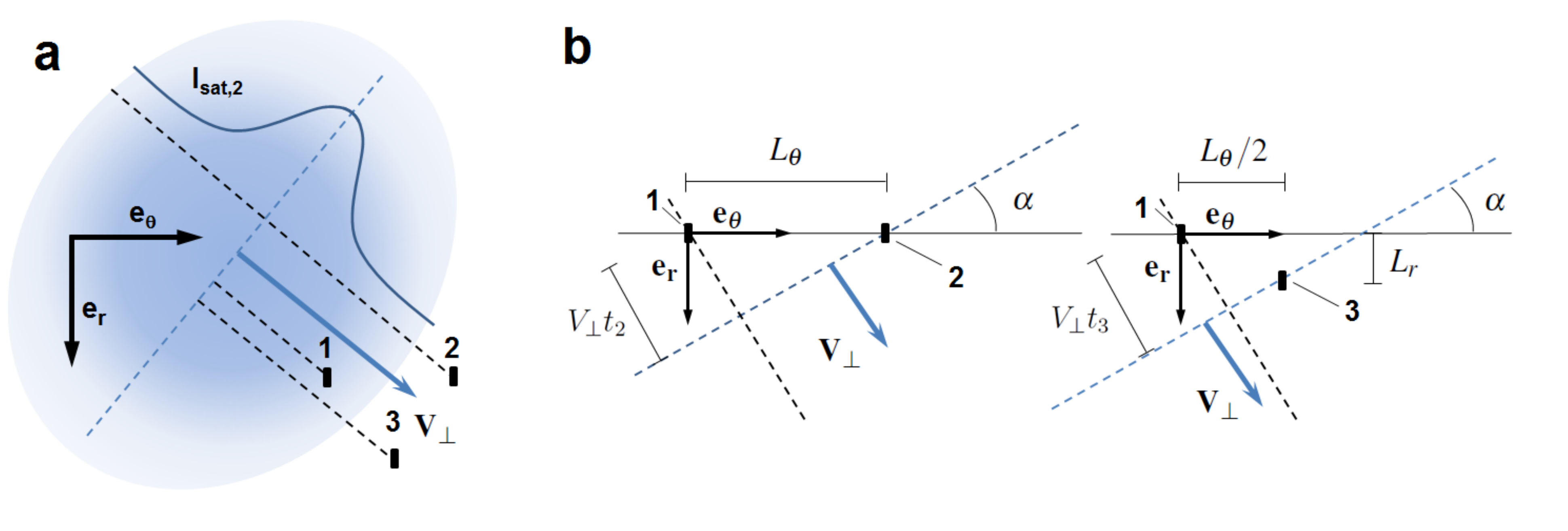}
	\caption{\textit{Filament propagation model. (a) A generic blob propagating towards the three $I_{sat}$ pins in Fig \ref{fig:diag}. $\bf{e}_r$ and $\bf{e}_\theta$ stand for radial and binormal directions. The blob measures 20 mm in the direction parallel to $V_\bot$ and the distance between pins is scaled accordingly. The relative trajectories of the pins are represented as dashed black lines. As an example, the signal measured by pin 2, $I_{sat,2}$ is represented over the trajectory. (b) Geometric description of the passage over the pins: $L_r$ and $L_\theta$ stand for the radial and binormal separation of the pins, $V_\bot$ is the perpendicular propagation velocity of the front/blob and $t_2$, $t_3$ are the correlation delays between pin 1 and pins 2 and 3.}}
	\label{fig_Ap}
\end{figure}

If these assumptions hold, the points of the blob in which the maximum level of signal are detected by each probe will be roughly distributed along a straight line. This can be seen in the cartoon in figure \ref{fig_Ap}a: as the blob propagates over the pins, the detection (maximum in the $I_{sat}$ signal happens at different times but always along the same dashed blue line traveling with it. In this case, the blob can be modeled as a front propagating at a velocity $V_\bot$ with an angle $\alpha$ with respect to the binormal direction. Both $V_\bot$ and $\alpha$ can be calculated from the cross-correlation time between pins 1 and 2, $t_2$, and 1 and 3,$t_3$.  Indeed, combining the geometric information of figure \ref{fig_Ap}b,

\begin{equation}
	V_\bot =\biggl( \biggl[\frac{t_2}{L_\theta}\biggr]^2+\biggl[\frac{2 t_3- t_2}{2 L_r}\biggr]^2 \biggr)^{-\frac{1}{2}}, \quad \sin{\alpha} = \frac{t_2}{L_\theta} V_\bot.
\end{equation}

From them, the obtainment of radial and binormal propagation velocities is straight forward: $V_\theta = V_\bot\sin{\alpha}$ and $V_r = V_{\bot} \cos{\alpha}$.\\

Note that the purpose of this model is not to provide a precise description of the blob structure, but to find some working hypothesis which allow to get a correct qualitative measure of the blob propagation: The first assumption is commonplace in filament theory, and is in good agreement with general blob observations \cite{Zweben07} and also with GEMR simulations of AUG SOL turbulence \cite{NoldPhd}. The second assumption is introduced a posteriori as a result of the initial results of the conventional analysis. In any case, the results of this model are in good agreement with it: as can be seen in Fig. \ref{fig:fig2}, the minimum size of the blob is around $\delta_b \simeq 5 mm$, with most blobs being substantially larger than that. As a reference, the blob represented in the cartoon in figure \ref{fig_Ap}a has a size of $\delta \simeq 20$ mm. Finally, the third assumption comes from the general notion of the blob being propelled by the $E\times B$ force caused by the dipole formed by the emergence of curvature drifts on the density fluctuation \cite{Krash01} (so in principle the axis of the roughly symmetric dipole should be perpendicular to $V_\bot$). Notice that no assumption is made on the size of the blob on the direction perpendicular to $V_\bot$. The reason for this is an intrinsic feature of Langmuir probe measurements (and not a result of this model): regardless of the blob shape and tilt, a single pin can only collect information about the evolution of the blob in the direction of its motion. Only through a complex reconstruction involving many pins can the shape of the blob in the direction perpendicular to it be assessed. As a result, not much can be said about the shape or orientation of the detected filaments: the could either be circular, tilted ellipses or structures elongated along the radial direction (streamers). In any case, the results seem to be reasonable when compared to other AUG diagnostics (see, eg., Fig \ref{fig:resumen}) or, as discussed in section \ref{Discussion}, with normalized blob measurements from other machines.\\

\end{document}